\documentclass{article}

\usepackage{braket}
\usepackage{float}
\usepackage{amsmath}
\usepackage{amssymb}
\usepackage{graphicx}
\usepackage{varioref}
\usepackage{hyperref}

\usepackage[a4paper, total={6in, 8in}]{geometry}

\newtheorem{defi}{Definition}
\newtheorem{example}{Example}
\newcommand{\intI}{\iota^\mathrm{I}}
\newcommand{\intII}{\iota^\mathrm{I\times U}}
\newcommand{\intIII}{\iota^\mathrm{N\times U}}
\newcommand{\evoI}{\epsilon^\mathrm{I}}
\newcommand{\evoII}{\epsilon^\mathrm{N}}

\begin{document}

\title{A Mathematical Definition of Particle Methods}

\author{
Johannes Bamme
\and
Ivo F.~Sbalzarini
}

\date{%
\footnotesize{
Technische Universit\"{a}t Dresden, Faculty of Computer Science, Dresden, Germany. \\
Max Planck Institute of Molecular Cell Biology and Genetics, Dresden, Germany. \\
Center for Systems Biology Dresden, Dresden Germany. \\
Cluster of Excellence Physics of Life, TU Dresden, Dresden, Germany. \\
Center for Scalable Data Analytics and Artificial Intelligence (ScaDS.AI) Dresden/Leipzig, Germany.
}
}
\maketitle

\begin{abstract}
We provide a formal definition for a class of algorithms known as ``particle methods''. Particle methods are used in scientific computing. They include popular simulation methods, such as Discrete Element Methods (DEM), Molecular Dynamics (MD), Particle Strength Exchange (PSE), and Smoothed Particle Hydrodynamics (SPH), but also particle-based image processing methods, point-based computer graphics, and computational optimization algorithms using point samples. All of these rest on a common concept, which we here formally define. The presented definition of particle methods makes it possible to distinguish what formally constitutes a particle method, and what not. It also enables us to define different sub-classes of particle methods that differ with respect to their computational complexity and power. Our definition is purely formal, independent of any application. After stating the definition, we therefore illustrate how several well-known particle methods can be formalized in our framework, and we show how the formal definition can be used to formulate novel particle methods for non-canonical problems.
\end{abstract}

\section{Introduction}
Particle Methods are a classic and popular class of algorithms in scientific computing, from applications such as computational plasma physics \cite{Hockney:1966} to computational fluid dynamics \cite{Cottet:1990}. Some of the historically first computer simulations in these domains were done using particle methods \cite{Verlet:1967, Chorin:1973} and the field is still under active development today \cite{cottet:2020,Caprace:2020}. A key advantage of particle methods is their versatility, as they can simulate both discrete and continuous models either stochastically or deterministically. 

In simulations of discrete models, particles naturally represent the discrete entities of the model, such as atoms in molecular dynamics simulations \cite{Alder:1957}, cars in simulations of road traffic \cite{SUMO2018}, or grains of sand in discrete-element simulations of granular flows \cite{Walther:2009}. When simulating continuous models, or numerically solving differential equations, particles represent mathematical collocation points or Lagrangian tracer points of the discretization of the continuous fields \cite{Reboux:2012,Cottet:2014}. The evaluation of differential operators on these fields can directly be approximated on the particles using numerical methods such as Smoothed Particle Hydrodynamics (SPH) \cite{Gingold:1977,Monaghan:2005}, Reproducing Kernel Particle Methods (RKPM) \cite{Liu:1995}, Particle Strength Exchange (PSE) \cite{Degond:1989a, Eldredge:2002}, or Discretization-Corrected PSE (DC-PSE) \cite{Schrader:2010,Bourantas:2016}. Also simulations of hybrid discrete-continuous models are possible, as often done in plasma physics, where discrete point charges are coupled with continuous electric and magnetic fields \cite{Hockney:1966}. In addition to their versatility, particle methods can also efficiently be parallelized on shared- and distributed-memory computers \cite{Karol:2018, Incardona:2019,Sbalzarini:2006b,Iwasawa:2016,Reynders:1996a}, and they simplify simulations in complex \cite{Sbalzarini:2005} and time-varying \cite{Bergdorf:2010} geometries, as no computational mesh needs to be generated and maintained. Particle methods have therefore been widely used in numerical simulations.

Beyond the field of simulations, however, structurally similar algorithms have been developed. Examples include particle-based image processing methods \cite{Cardinale:2012, Afshar:2016}, point-based computer graphics \cite{Gross:2011uz}, and computational optimization algorithms using point samples, e.g.~\cite{Hansen:1996, Muller:2010}. The computational and algorithmic structure of 
these methods is analogous to that of particle methods traditionally used in computer simulations. 

Due to this similarity, we here propose a broader definition of particle methods to include all computational algorithms that use points (i.e., particles or  zero-dimensional elements) with positions in some space (e.g., physical space, phase space, parameter space) and that carry some properties. These points can evolve their positions and properties as a result of interacting with each other. Therefore, particles are zero-dimensional computational objects that store positions and properties and are able to {\em interact} and {\em evolve}. We treat the position of a particle as one of its properties, even though in practical applications and implementations the position property is often specifically distinguished. We also restrict particles to only interact pairwise. Higher-order interactions are not permitted and must be reduced to sequences of pairwise interactions using additional particle properties. 
This leads to a generalized definition of particle methods across application domains, from simulation to graphics, to computer vision and optimization.
In this generalized view, of course, the question arises what formally constitutes a particle method and, in reverse, what is not a particle method.

We address this fundamental question by: (1) providing a formal definition of particle methods independent of any application, and (2) illustrate how several classic and well-known particle methods can be formalized in our framework. Illustrating the value of a generic formal definition, we further show that particle methods---according to our definition---can be written for solving problems that were not so far considered typical applications of particle methods. 

The present formal definition establishes particle methods as an algorithmic class with a well-defined structure. As a consequence, it becomes possible to write novel particle methods for many problems, guided by the scaffold provided by the definition. This, however, makes no statement about the computational complexity or efficiency of such a method. In many cases, particle methods are not expected to be particularly efficient solutions. They do, however, provide an appealing, since unifying, algorithmic framework with a structure and regularity that can be exploited, for example, in software engineering of scientific codes or in the design of programming languages for scientific computing.

\section{Nomenclature and Notation}\label{sec:def}

Before stating our formal definition of particle methods, we introduce the notation and nomenclature used and define the underlying mathematical concepts. 

\begin{defi} \label{def:def1}
The \textbf{set of all tuples $A^*$} from the elements of a set $A$, including the empty tuple $()$, is defined using the Cartesian product as follows:
\begin{align}
	A^0:=& \{ () \} \\
	A^1:=& A \\
	A^{n+1}:=& A^n\times A \quad \text{ for } n \in \mathbb N_{>0}\\
	A^*:= & \bigcup_{j=0}^{\infty} A^j \qquad \Big( = \{()\} \;\;\cup\;\; A \;\;\cup\;\; A\times A \;\;\cup\;\; A\times A \times A \;\;\cup\;\; ... \;\;\Big).
\end{align}
\end{defi}

We use the following notational conventions:
\begin{align}
	\mathbf p &\in P^* && \text{bold symbols for tuples of arbitrary length }\\
	\mathbf p &= (p_1,...,p_n)&& \text{regular symbols with subscript indices for the elements of these tuples}\\
	\vert \mathbf p \vert &:= n && \text{a tuple in vertical bars is the length of the tuple }\\
	 p &= (a,b,c) && \text{a tuple prototype of determined length with specific element names}\\
	 p_j &= (a_j,b_j,c_j)  && \text{an indexed tuple of determined length with named elements}
\end{align}

We further use:
\begin{defi} \label{def:compo}
The \textbf{composition operator} $*_h$ of a binary function $h$: Be $h: A \times B \rightarrow A$ then $*_h: A\times B^*  \rightarrow A$ is recursively defined as:
\begin{align}
	&a *_h () := a\\
	&a *_h (b_1,b_2,...,b_n) := h(a,b_1) *_h (b_2,...,b_n).
\end{align}
\end{defi}

\begin{example}
For $-$ the ordinary arithmetic subtraction, the operator $*_-$ is:
\begin{align*}
	&9 *_- () = 9,\\
	&13 *_- (3,4,1) = (13-3) *_- (4,1) = ((13-3)-4)-1 = 5.
\end{align*}
\end{example}

\begin{defi} \label{def:concat}
The \textbf{concatenation} $\circ : A^* \times A^* \rightarrow A^*$ of tuples $\mathbf a, \mathbf b\in A^*$ is defined as:
\begin{align}
	\mathbf a \circ \mathbf b =& \left(a_1,...,a_n\right) \circ \left(b_1,..,b_m \right) \\
	:= & \left(a_1,...,a_n,b_1,..,b_m \right). \notag
\end{align}
\end{defi}

Finally, we define a notational convention for a subtuple $\mathbf{ b} \in A^*$ of a tuple $\mathbf a\in A^*$:

\begin{defi} \label{def:subtuple}
Be $f: A^* \times \mathbb N \rightarrow \{\top, \bot \} $ ($\top = true$, $\bot = false$) the condition for an element $a_j$ of the tuple $\mathbf a$ to be in $\mathbf b$, then $\mathbf b:=(a_j \in \mathbf a: f(\mathbf a, j))$ defines a subtuple of $\mathbf a$ as follows:
\begin{align}
	&\mathbf{ b} := (a_{j_1},...,a_{j_n}) = (a_j \in \mathbf a: f(\mathbf a, j))  \\
	 \Leftrightarrow \quad& \mathbf a= (a_1,..., a_{j_1},...,a_{j_2},...,a_{j_n},...,a_m) \quad \land \quad \forall k\in \{1,..,n\}:  \; f(\mathbf a, j_k)=\top . \notag
\end{align}
\end{defi}

\begin{example}
Be $\mathbf{\overline a}= (4,1,1,5,66,3,4,30)$ and $f(\mathbf a, j):=(a_j < 5)$ then
\begin{align*}
	\mathbf{\overline b} :=&(\overline a_j \in \mathbf{\overline a} : f( \mathbf{\overline a}, j))\\
	 =& (\overline a_j  \in (4,1,1,5,66,3,4,30) : \overline a_j< 5)\\
	 =& (4,1,1,3,4).
\end{align*}
\end{example}

An \textbf{index tuple} $\mathbf c \in \mathbb N^*$ can be denoted in a similar way: Be $f: \mathbb N \rightarrow \{\top, \bot \} $ then $\mathbf c:=(j \in \mathbb N: f(j))$ defines an index tuple as follows:
\begin{align}
	&\mathbf{ c}  :=  (j_1,..., j_n) = (j \in \mathbb N: f(j)) \\
	\Leftrightarrow \quad& j_1 < j_2 < ... < j_n \quad \land \quad \forall k\in \{1,..,n\}: \; f(j_k)=\top . \notag
\end{align}

\section{Illustrative Example} \label{sec:defPM:expample} \label{sec:defPM:DEM:PM}
In order to illustrate our definition of particle methods, we use a single, simple example thoughout this manuscript. 
This example is motivated by the Discrete Element Method (DEM)~\cite{Silbert:2001}, but is greatly simplified and reduced to its key constituents. This purely serves didactic purposes and has no ambition to resemble any real physics, but only to illustrate our abstract concepts in a simple and concrete case.

The example we consider models perfectly elastic collisions between spheres of uniform diameter $d$ and constant unit mass in a one-dimensional continuous space.
The space is without boundaries, such that no boundary conditions are required.

In this example, position $x$ of an individual sphere changes over time $t$, as:
\begin{align}
 &x(t)=x_0+ \int_0^t v(t) \,\mathrm{d}t ,
\end{align}
where $v$ is the velocity of the sphere and $x_0$ its initial position. 
The explicit Euler time-stepping algorithm discretizes this equation in time using a fixed time step size $\Delta t$, yielding:
\begin{align}
\label{eq:exp:euler}
 &x^{t+\Delta t}=x^{t}+  v(t)\Delta t,
\end{align}
which is iterated until a given final time $T$.
A perfectly elastic collision between two spheres $1$ and $2$ results in them swapping their velocities.
The positions before the collision $x^n$ and the positions after the collision $x^{n+1}$ remain the same, leading to the following collision rules:
\begin{align}
\label{eq:exp:collision1}
 x^{n+1}_1 &=x_1^{n},\\
 x^{n+1}_2 &=x_2^{n},\\
 v^{n+1}_1 &=v_2^{n},\\
 \label{eq:exp:collision4}
 v^{n+1}_2 &=v_1^{n}.
\end{align}
Two spheres $1$ and $2$ are considered colliding if and only if $\vert x_2-x_1\vert \leq d$.

\section{Definition of Particle Methods}\label{sec:defPM} 

We state a formal definition of the algorithmic class of particle methods using the mathematical concepts from Section \ref{sec:def}, and we illustrate every line of the definition using the example from Section \ref{sec:defPM:expample}. 

Certainly, our definition is not unique. There are alternative ways of formulating the same concepts. We chose the specific formulation here because it naturally relates to the structural elements of a practical implementation, i.e., to the methods and subroutines often found in particle methods software. Our definition is based on observing and analyzing a large number of existing particle methods and distilling their structural commonalities. We state the definition in its most general form, realizing that many practical examples of particle methods do not require the full expressiveness. This can then later be used to define sub-classes of particle methods.

We subdivide the definition of particles into three parts: First, the definition of the data structures and functions. These data structures and functions are the structural components of algorithms written as particle methods. Second, the definition of the instance of a particle method. The instance of a particle method describes a specific problem, which can then be solved or simulated using the data structures and functions of the particle method. Third, the definition of the state transitions. The state transitions describes how an instance of a particle method proceeds from one state to the next by using its data structures and functions. The transition functions can be different for different computer architectures and place additional constrains on the data structures and functions. We present them in the most general form without specifying any further limitations at this stage.

\subsection{Data structures and functions}\label{sec:defPM:definition} 
The definition of a particle method encapsulates the structural elements of its implementation in a small set of data structures and functions that need to be defined at the onset. 
This follows a similar logic as some definitions of Turing machines~\cite{Kozen:1997}. Both concepts are state transition systems working on discrete objects. 
Specifically, in our definition, a particle methods needs to define two data structures and five functions as follows:

\begin{defi}
A \textbf{particle method} is a 7-tuple $(P, G, u, f, i, e, \mathring e)$, consisting of the two data structures
\begin{align}
    \label{eq:defPM:P}
    &P  := A_1 \times A_2 \times ... \times A_n 
        &&\text{the particle space,}\\
    \label{eq:defPM:G}
    &G := B_1 \times B_2 \times ... \times B_m  
        &&\text{the global variable space,}
\end{align}
such that $[G\times P^*]$ is the {\em state space} of the particle method, and five functions: 
\begin{align}
    \label{eq:defPM:u}
    &u: [G \times P^*] \times \mathbb N \rightarrow \mathbb N^*
        &&\text{the neighborhood function,}\\
    \label{eq:defPM:f}
    &f:  G \rightarrow \{ \top,\bot \} 
        &&\text{the stopping condition,}\\
    \label{eq:defPM:i}
    &i:  G \times P \times P \rightarrow P\times P  
        &&\text{the interact  function,}\\
    \label{eq:defPM:e}
    &e:  G \times P\rightarrow G \times P^*  
        &&\text{the evolve function,}\\
    \label{eq:defPM:eg}
    &\mathring{e} :  G \rightarrow G   
        &&\text{the evolve function of the global variable.}
\end{align}
\end{defi}

These are the only objects that need to be defined by the user in order to specify a particle method. In the remainder of this section, we provide additional explanations, examples, and clarification to help understand the above formal definitions.

\subsubsection*{Explanation of the Particle Space $P$ (eq. \ref{eq:defPM:P})}
\begin{align}
        P:=A_1 \times ... \times A_n \tag{eq. \ref{eq:defPM:P}}
\end{align}
The structure of the particle space $P$ follows directly from the observation that in any particle method, the particles are points in some space that carry/store some properties.
Hence, a particle in our definition is a tuple of properties, with its position being one of them.
The particle space is the Cartesian product of the  property sets $A_1,..., A_n$, since every of the $n$ properties of a particle can be of a different data type and hence live in a different space.
Examples of particle properties can include, depending on the application, a color, a velocity, an acceleration, a Boolean flag, a position in some space, a vector, a name,  and so on. 
\\[12pt]
In the example from Section \ref{sec:defPM:expample}:
\begin{align}
     P&:= \mathbb R \times \mathbb R \hspace{2cm} (n=2, \quad A_1, A_2 = \mathbb R)\\
     p&:=(x,v) \in P
\end{align}
In the example, particles are spheres of identical size and mass.
A particle is therefore fully described by its position $x$ and its velocity $v$.
This means that it is reasonable to choose the space of particles as $P:= \mathbb R \times \mathbb R$.

\subsubsection*{Explanation of the Global Variable Space $G$ (eq. \ref{eq:defPM:G})}
\begin{equation}
    G:=B_1 \times ... \times B_m \tag{eq. \ref{eq:defPM:G}}
\end{equation}
The structure of the global variable space $G$ in analogous to that of the particle space.
A global variable is also a tuple of properties.
The global variable is not necessary from the perspective of  computational power, but it simplifies the formulation of many particle methods by encapsulating simulation properties that are not specific to a particle.  
This improves readability and possibly implementation efficiency.
The global variable is accessible throughout the particle method, such that all functions can depend on it.
The global variable space is the Cartesian product of the arbitrary property sets $B_1,..., B_m$.
In general, the property sets $B_1,..., B_m$ of the $m$ components of the global variable can be of arbitrary  data type. Examples of global properties include the time step size of a simulation, the total number of particles in the simulation, the total energy of the system, and so on.
\\[12pt]
In the example from Section \ref{sec:defPM:expample}:
\begin{align}
	G&:= \mathbb R \times \mathbb R \times \mathbb R\times \mathbb R \hspace{1.5cm} (m=4, \quad B_1,..., B_4 = \mathbb R) \\
	g&:= (d, t, \Delta t, T) \in G \
\end{align}
In the example, the global variable $g:=( d, t, \Delta t, T) \in G$ is the collection of the non-particle-specific diameter $d$ of the spheres and the time parameters of the simulation, the current time $t$, the simulation time step size $\Delta t$, and the stopping time $T$. Again, all can be chosen as real numbers. If instead of storing the time, one would store the index of the current time step, then the corresponding set could be $\mathbb{N}$ instead of $\mathbb{R}$.

\subsubsection*{Explanation of the State Space $[G \times P^*]$}
\begin{equation}
    [G\times P^*]  \notag
\end{equation}
The state of a particle method collects all information about a particle method at a certain time point.
Hence, the state of a particle method consists of a global variable and potentially many particles collected in a tuple of particles.

Formally, the state space of the particle method is the Cartesian product of the global variable space $G$ and the set of all tuples of particles $P^*$ from the particle space $P$ (see definition \ref{def:def1}). An element $[g^t, \mathbf p^t] \in [G \times P^*]$ fully describes the state of a particle method at a time point $t$. We use the square brackets ($[ \cdot  ]$) to mean that the state is one element even thought we write the global variable $g$ and the particle tuple $\mathbf p$ separately.

\subsubsection*{Explanation of the Neighborhood Function $u$ (eq. \ref{eq:defPM:u})}
\begin{equation}
    u: [G \times P^*] \times \mathbb N \rightarrow \mathbb N^*  
    \tag{eq. \ref{eq:defPM:u}}
\end{equation}
The neighborhood function $u$ is the only function that, in general, depends on the entire state of the particle method $[g,\mathbf p]$.
We introduce this function to reduce computation.
Without it, each particle would need to interact with every other particle in order to decide if they are contributing to their respective changes.
If the number of contributing interactions is low compared to the number of all particles, then the neighborhood function helps reduce the computation by returning the indices of only those interaction partners that actually contribute to the result.
For instance, the neighborhood can be defined by a cut-off radius beyond which particles no longer ``feel'' each other. Cell list~\cite{Quentrec:1973} or Verlet list~\cite{Verlet:1967} can then be used to implement a neighborhood function with linear run-time. In general, however, a neighborhood need not be defined geometrically, but can be an arbitrary set of particle indices. 

The neighborhood function therefore operates on particle indices.
It takes an index as input, pointing to a particle in the current state, and returns a tuple of indices, pointing to the neighboring particles of the input particles.
We chose indices instead of particles directly because an index is a unique identifier of an element in a tuple.
Besides, indices provide stable identifiers of particles throughout the whole interaction phase.
They remain the same while the particle itself may change.

Formally, the neighborhood function maps a state of a particle method and an index of a particle to a tuple of particle indices. It returns the indices of all interaction partners of the particle with input index.  
\\[12pt]
In the example from Section \ref{sec:defPM:expample}:
\begin{align}
     \label{eq:expPM:u}
   u([g, \mathbf p], j)  := ( k\in \mathbb N: p_k, p_j\in\mathbf p \;\land\;  0 < x_k - x_j  \leq d)
\end{align}
In the example, the neighborhood function $u$ returns an index tuple (def. \ref{def:subtuple}) of all collision partners of the particle $p_j \in \mathbf p$.
In this case, the neighborhood function is written such that each collision pair is considered only once, i.e., if $p_k$ is in the neighborhood of $p_j$ than $p_j$ shall not be in the neighborhood of $p_k$. This is a matter of definition and typically known as ``asymmetric neighborhood'' in particle methods. 
If the neighborhood function $u$ would be: $u([g, \mathbf p], j):=(k\in \mathbb N: \; p_k, p_j\in\mathbf p \;\land\; k\neq j  \;\land\;  | x_k - x_j | \leq d)$, then both collision partners would show up as neighbors of the respective other.

\subsubsection*{Explanation of the Stopping Condition $f$ (eq. \ref{eq:defPM:f})}
\begin{equation}
    f:  G \rightarrow \{ \top,\bot \}  \tag{eq. \ref{eq:defPM:f}}
\end{equation}
The stopping condition $f$ only depends on the global variable. 
This choice is not limiting, because the global variable can be changed by every particle at any time.
This means that each particle can influence the outcome of the stopping condition.

Formally, the stopping condition is a function that maps a global variable to a Boolean value, $\top$ (true) or $\bot$ (false). It determines when the algorithm ends. It ends when $f(g)~=~\top$.
\\[12pt]
In the example from Section \ref{sec:defPM:expample}:
\begin{equation}
 \label{eq:expPM:f}
    f(g) := (t \geq T)
\end{equation}
In the example, the algorithm terminates when the current time $t$ reaches or exceeds the stopping time $T$, then $f(g)~=~\top$.

\subsubsection*{Explanation of the Interact Function $i$ (eq. \ref{eq:defPM:i})}
\begin{equation}
    i:  G \times P \times P \rightarrow P\times P  \tag{eq. \ref{eq:defPM:i}}
\end{equation}
A particle methods proceeds by computing interactions between particles. The basic building block for this is the interact function, which specifies how two particles interact with each other.
Higher-order interactions, e.g. three-body interactions, are realized by a series of pairwise interactions.
The interact function can change both particles.
This provides a performance advantage in cases of symmetric interactions.
It can therefore not only read information from the interacting particles, but also write information into the interacting particles.

Formally, the interact function maps a triple of a global variable and two particles to two particles. It describes the changes the interaction causes to the properties of both interacting particles. While in general, $i$ can change both particles, the change can be zero for one or both of them. The interact function is only applied for pairs of particles identified by the tuple of indices provided by the neighborhood function $u$.
\\[12pt]
In the example from Section \ref{sec:defPM:expample}:
\begin{equation}
    \label{eq:expPM:i}
    i(g, p_j, p_k) := \left( (x_j, v_k),  (x_k, v_j) \right)
\end{equation}
In the example, an interaction amounts to an elastic collision between two spheres, as described by the collision rules (eqs. \ref{eq:exp:collision1} -- \ref{eq:exp:collision4}). Hence, the two involved particles $p_j$ and $p_k$ keep their positions $x_j$ and $x_k$, but swap their velocities $v_j$ and $v_k$.
In this case the interact function $i$ changes both particles. Therefore the neighborhood function $u$ defines an asymmetric neighborhood, which prevents duplicate  interactions. Meaning, if $p_k$ is in the neighborhood of $p_j$ than $p_j$ will not be in the neighborhood of $p_k$. Hence, they interact just once {\em as a pair}.
In practical applications, this type of symmetric interaction not only reduces the computational cost of the algorithm, but it also helps conserve quantities such as energy.
If the interaction would change only one particle, we would need an additional mechanism to ensure that no kinetic energy is lost, e.g. by using additional properties to keep a running tally of the net velocity exchanged.

\subsubsection*{Explanation of the Evolve Function $e$ (eq. \ref{eq:defPM:e})}
\begin{equation}
    e:  G \times P\rightarrow G \times P^* \tag{eq. \ref{eq:defPM:e}}
\end{equation}
The evolve function $e$ changes the properties of a particle due to its own properties and the global variable. This includes any change that is independent of interactions with other particles.
This provides a place to update properties that need to stay constant during all interactions, or to reset properties that serve as temporary accumulators used during the interactions. It is also the place to implement autonomous dynamics, i.e., dynamics that for example only depends on time, but not on the properties of the other particles. Examples include simulations of the chemical reaction terms in a spatio-temporal reaction-diffusion simulation, or autonomous stochastic processes.

Formally, the evolve function maps a global variable and a particle to a global variable and a tuple of particles. It can therefore change the global variable, for instance to implement global reduction operations like summing a property over all particles. Since the result is a tuple of particles, $e$ can also create or destroy particles. This is for example used in population dynamics simulations or in adaptive-resolution methods for continuous models \cite{Reboux:2012}. Hence, $e$ potentially changes the total number of particles~$|\mathbf{p}|$.
\\[12pt]
In the example from Section \ref{sec:defPM:expample}:
\begin{equation}
    \label{eq:expPM:e}
    e(g,p_j) := \left( g, \; \big(( x_j + \Delta t \cdot v_j, \; v_j )\big)\right)
\end{equation}
In the example, the global variable $g$ remains unchanged by the evolve function, even though this is not necessary as per the definition.
The position $x_j$ of the particle $p_j$ in the example is evolved according to the velocity $v_j$ and the time step size $\Delta t \in g$ using explicit Euler time-stepping (eq. \ref{eq:exp:euler}).
The velocity $v_j$ of the particle $p_j$ stays the same.

\subsubsection*{Explanation of the Evolve Function of the Global Variable $\mathring e$ (eq. \ref{eq:defPM:eg})}
\begin{equation}
    \mathring{e} :  G \rightarrow G  \tag{eq. \ref{eq:defPM:eg}}
\end{equation}
The evolve function of the global variable $\mathring e$  changes the global variable based only on the current value of the global variable.
We chose to have this function because otherwise the only  place to change the global variable would be the evolve function of the particles, which would make it difficult to prevent multiple evolutions of the global variable.
Hence, the evolve function of the global variable provides us with a structure to describe singleton global behavior of the simulation, such as for example increasing the simulation time after completion of a time step.

Formally, the evolve function of the global variable maps a global variable to a global variable. It describes the changes to the global variable due to its own values. This allows updating the global variable without requiring a global operation.
\\[12pt]
In the example from Section \ref{sec:defPM:expample}:
\begin{equation}
    \label{eq:expPM:eg}
    \mathring e(g) := (d, t+\Delta t, \Delta t, T)
\end{equation}
In the example, the evolve function of the global variable $\mathring e$, increments the current time $t$ by the time step size $\Delta t$.
All other global properties remain unchanged.

\subsection{Particle method instance} \label{sec:defPM:instance}
Using the above definitions of data structures and functions of a particle method, we define an {\em instance} of a particle method as a specific realization. This requires specifying the initial set of particles and the initial value of the global variable, as follows:

\begin{defi} \label{def:PMI} 
An \textbf{instance} of a particle method $(P, G, u, f, i, e, \mathring e)$ is defined by an initial state:
\begin{equation}
    \label{eq:defPMI:gp}[g^1,\mathbf{p}^1] \in [G\times P^*]
\end{equation}
 consisting of initial values for the global variable $g^1 \in G$ and an initial tuple of particles $\mathbf p^1 \in P^*$.
\end{defi}


The particle method instance specifies the starting point of a particle method. We chose to separate the definition of the instance of a particle method from the definition of the particle method itself in order to distinguish the specific problem from the general algorithm. 

Formally, a particle method instance $[ g^1,\mathbf{p}^1 ]$ is an element of the state space $[G\times P^*]$ of a particle method. Since an element $[g^t, \mathbf p^t] \in [G \times P^*]$ fully describes the state of a particle method at a time point $t$, the particle method instance $[g^1,\mathbf{p}^1]$ fully describes the state of a particle method.
\\[12pt]
In the example from Section \ref{sec:defPM:expample}:
\begin{align}
	\label{eq:expPMI:gp}
	[g^1,\mathbf p^1] & \in [G\times P^*] &&\text{initial state of the particle method}\\
	\notag \\
	\label{eq:expPMI:g}
	g^1 :=& (d, t, \Delta t, T) \in G &&\text{initial global variable}\\
	d&:= 0.5 &&\text{diameter of the spheres}\notag\\
	t&:= 0   &&\text{initial time}\notag\\
	\Delta t&:= 0.1&&\text{time step size}\notag\\
	T&:= 10&&        \text{stopping time}\notag\\
	\notag\\
	\label{eq:expPMI:p}
	\mathbf p^1 :=& \big( p_1 ,p_2 ,p_3 \big) \in P^* &&\text{initial particle tuple}\\
	p_1&:=(0,2)&&\text{particle 1 with $x_1=0$ and $v_1=2$}\notag\\
	p_2&:=(0.49, -1)&&\text{particle 2 with $x_2=0.49$ and $v_2=-1$}\notag\\
	p_3&:=(2,1)&&\text{particle 3 with $x_3=2$ and $v_3=1$}\notag
\end{align}
This example of an particle method instance is the initial sate  for the didactic example from Section \ref{sec:defPM:expample}.
It is a 1D elastic collision simulation of three spheres with the same diameter and mass. 
Since mass cancels out if it is the same for all spheres we do not explicitly store it.
This particle method instance defines the initial state $[g^1, \mathbf p^1]$. The initial values of the properties of the global variable $g^1$ are the diameter of the spheres $d=0.5$, the initial time $t=0$, the time step size $\Delta t= 0.1$, and the stopping time $T=10$. The initial particle tuple $\mathbf p$ consists of 3 particles, representing three spheres. These particles are $p_1=(0,2)$, $p_2=(0.49, -1)$, and $p_3=(2,1)$. As one can see, $p_1$ and $p_2$ are all ready colliding. Their distance is smaller than $d$, and their velocities point toward each other.

\subsection{State transition} \label{sec:defPM:transition}
In a specific particle method, the elements of the tuple $(P, G, u, f, i, e, \mathring e)$ (eqs. \ref{eq:defPM:P} -- \ref{eq:defPM:eg}) need to be specified. Given a certain starting point as defined by an instance, the algorithm then proceeds in iterations. Each iteration corresponds to one state transition that advances the current state of the particle method $[g^{t},\mathbf{p}^{t}]$ to the next state $[g^{t+1},\mathbf{p}^{t+1}]$, starting at the initial state $[g^1,\mathbf p^1]$.
The state transition uses the functions $u, f, i, e, \mathring e$ to determine the next state in a canonical way. It is hence the same for every particle method and does not need to be defined by the user. We formally define:
\begin{defi}\label{def:PMS}
    The \textbf{particle method state transition function} $s$ is a partial function mapping one state to another state:
    \begin{equation}
    \label{eq:defPM:s}
        s :  [G\times P^*] \nrightarrow [G\times P^*].
    \end{equation}
    Iterating this generates a sequence of states. The sequence ends when $s$ is undefined, i.e.:
    \begin{equation}
    \label{eq:defPMS:lastS}
    \begin{split}
        &[ g^T, \mathbf p^T] \text{ is the last state of the state sequence} \\
        \Longleftrightarrow \quad&s\left([ g^T, \mathbf p^T]\right)= undef \; \land\; \forall t\in \{1,...,T-1\}: s\left([ g^t, \mathbf p^t]\right)\neq undef
    \end{split}
    \end{equation}
\end{defi}

While the state transition function $s$ is the same for all particle methods, its definition and implementation can be different for different computer architectures. Frequent constraints in practical applications include indexing-order independence on multi-threaded architectures, interaction order limitations on distributed-memory architectures, or data writing synchronizaton constraints on graphics cards. 

We here present the most generic form of the state transition function on a sequential computer. The only assumption we use is that the particle interactions are evaluated sequentially one after another. Further limitations can be imposed if needed, leading to more specific state transition functions. 
    
We define the general sequential state transition function $s_\mathrm{general}$ using five sub-functions $\intI$, $\intII$, $\intIII$, $\evoI$, $\evoII$. In the most general case, all particles interact with all of their respective interaction partners (possibly all particles), as given by the neighborhood function $u$. This can be broken down into a loop over all particles ($\intIII$), where each particle interacts with its specific neighbors ($\intII$). Each of these interactions, in turn, is a pairwise interaction between two particles ($\intI$ as defined by $i$). The evolution happens after all interactions are completed. It can be broken down again into a loop over all particles ($\evoII$), where each individual particle evolves ($\evoI$).
    
After each interaction, in our definition, the current particle tuple is updated, such that the result of the interaction can affect the following interactions.
The same is true for the evolve function concerning the global variable $g$.
While this is not required for many practical examples, we include this possibility in the definition for generality.
In general, therefore, the result of the algorithm depends on the order in which the interactions and evolutions are evaluated, and hence on the index ordering of the particles.
The result of the algorithm becomes independent of particle ordering if the interaction results, as well as the global variable during the evolution, are reduced by a commutative operation, usually the addition, and accumulated in a separate particle property until the final evolve step $\evoII$.

The functions $s_\mathrm{general}, \intI, \intII, \intIII, \evoI, \evoII$ recursively call each other and are based on the particle method definition functions $u, f, i, e, \mathring e$.
The function $\intI$ builds upon the interact function $i$, the function $\intII$ upon $\intI$ and $u$, the function $\intIII$ upon $\intII$, the function $\evoI$ upon $e$, the function $\evoII$ upon $\evoI$, and finally the state transition function $s_\mathrm{general}$ brings it all together by building upon $\intIII$, $\evoII$, $\mathring e$ and $f$. This defines the following recursions:
\begin{align}
        \label{eq:defPM:sgeneral}
            s_\mathrm{general} &:  [G\times P^*] \nrightarrow [G\times P^*] 
            &&\text{sequential particle method state transition function,}\\[12pt]
        \label{eq:defPM:i1}
            \intI&: [G \times P^*] \times \mathbb N \times \mathbb N \rightarrow P^*
            &&\text{interaction  of  two  particles,}\\
        \label{eq:defPM:i2}
            \intII&: [G \times P^*] \times \mathbb N \rightarrow P^*
            &&\text{interaction of a particle with all its neighbors,}\\
        \label{eq:defPM:i3}
            \intIII&: [G \times P^*] \rightarrow P^*
            &&\text{interactions of all particles with all their neighbors,}\\
        \label{eq:defPM:e1}
            \evoI&:  [G \times P^*] \times P^* \times \mathbb N \rightarrow [G \times P^*]
            &&\text{evolution of one particle,}\\
        \label{eq:defPM:e2}
            \evoII&:  [G \times P^*]  \rightarrow [G \times P^*] 
            &&\text{evolution of all particles.}
\end{align}
Using the above formal definition of a particle method, all of these functions are completely defined and can be formally written down.
For this, all $g \in G$,  $\mathbf p \in P^*$, $[g,\mathbf p] \in [G\times P]$, and $j,k \in \mathbb N$. Over-bars indicate intermediate results that are used in the subsequent sub-functions, e.g. the result $\left(\overline p_j, \overline p_k\right)$ of the interact function $ i\left( g,p_j,p_k\right)$ is used in $\intI$. The composition operator $*_h$ (see Definition \ref{def:compo}) is only defined for functions $h$ with exactly two arguments. Therefore, functions with more than two arguments need to be rewritten in an indexed form. For example, the function $\intI ([g,\mathbf p], j,k)$ has 3 arguments and is hence rewritten as $\intI_{(g,j)}(\mathbf p,k)$, such that only the two arguments $\mathbf p$ and $k$ remain, whereas $g$ and $j$ became indices. This then yields the following formal expressions for the above recursive set of functions:
    \begin{align}
	\begin{split}
	\label{eq:defPM:si1}
	    \intI ([g,\mathbf p], j,k) &:=(p_1,..,p_{j-1},\overline p_j, p_{j+1},...,p_{k-1},\overline p_k, p_{k+1},...,p_{\vert\mathbf p\vert}),  \\
	    &\hspace{40mm} \text{with }\quad   (p_1,...,p_{\vert \mathbf p\vert}) = \mathbf p , \quad\left(\overline p_j, \overline p_k\right) := i\left( g,p_j,p_k\right),
	\end{split}
	\\[6pt]
	\label{eq:defPM:si2}
	    \intII ([g,\mathbf p], j) &:= \mathbf p \; *_{\intI_{(g,j)}} \; u([g,\mathbf p], j) 
	    \hspace{6.3mm}\text{with } \quad \intI_{(g,j)}(\mathbf p,k):=\intI ([g,\mathbf p], j,k), 
	\\[6pt]
	\label{eq:defPM:si3}
    	\intIII ([g,\mathbf p])&:=\mathbf p \; *_{\intII_g} (1,..,\vert \mathbf p\vert) \hspace{8.5mm} \text{with } \quad \intII_g (\mathbf p, j):=\intII ([g,\mathbf p], j), 
	\\[6pt]
	\label{eq:defPM:se1}
    	\evoI \big(  [g,\mathbf p], \mathbf q, j \big)&:= \big[ \overline g, \quad \mathbf q \circ \overline{\mathbf q}   \big]  
	    \hspace{15.5mm} \text{ with } \quad \left(\overline g, \overline{\mathbf q}\right) := e(g,p_j),
	\\[6pt]
	\label{eq:defPM:se2}
    	\evoII \big(  [g,\mathbf p]\big) &:= \big[ g, () \big] \; *_{\evoI_{\mathbf p}} \; (1,..,\vert \mathbf p\vert)   
    	\hspace{3mm} \text{with} \quad \evoI_{\mathbf p} \big( [g,\mathbf q], j \big)  := \evoI \big( [g,\mathbf p], \mathbf q, j \big),
	\\[12pt]
	\label{eq:defPM:ss}
    	s_\mathrm{general} \left(  [g, \mathbf p] \right)&:= 
    	\begin{cases}
    	\textit{undef} & \text{if} \qquad f\left(g\right)\\
        \big[ \mathring{e}( \overline{ g} ),\; \overline{\mathbf p}   \big]  & \text{else} 
       \end{cases} \hspace{1cm} \text{with } 
        \quad  \left[\overline g, \overline{\mathbf p}\right] := \evoII \big(  [g,\; \intIII ([g,\mathbf p])]\big),
    \\[6pt]
    \label{eq:defPM:tpp}
    [ g^{t+1}, \mathbf p^{t+1}] &:= s\left(  [g^{t}, \mathbf p^{t}]\right) .
\end{align}

For the following explanations, we use the example of the particle method definition (sec. \ref{sec:defPM:definition}) and the example of the particle method instance (sec. \ref{sec:defPM:instance}) for the illustrative example from Section \ref{sec:defPM:expample}.

\subsubsection*{Explanation of the First Interaction Sub-function $\intI$ (eqs. \ref{eq:defPM:i1}, \ref{eq:defPM:si1})}
\begin{align}
	\tag{eq. \ref{eq:defPM:i1}}
	 \intI &: [G \times P^*] \times \mathbb N \times \mathbb N \rightarrow P^*  \\
	 \tag{eq. \ref{eq:defPM:si1}}
	 \intI ([g,\mathbf p], j,k)&:=(p_1,..,p_{j-1},\overline p_j, p_{j+1},...,p_{k-1},\overline p_k, p_{k+1},...,p_{\vert\mathbf p\vert})  \\
	&\hspace{2cm} \text{with }\quad   (p_1,...,p_{\vert \mathbf p\vert}) = \mathbf p , \quad\left(\overline p_j, \overline p_k\right) := i\left( g,p_j,p_k\right)\notag 
\end{align}
The first interaction sub-function $\intI$ describes how the interaction of two particles contributes to the state change of the particle method. It is the core component of the loop over all particles and over all neighbors of each particle. The function $\intI$ evaluates the interaction function $i$ for a given pair of particles and a given current state of the particle method. It then takes the changed particles that $i$ returns and stores this change again in the state. We choose this encapsulation over directly evaluating $i$ because it requires no external storage to accumulate the changes of subsequent interactions.

Formally, this function maps a particle method state and two particle indices to a tuple of particles.
The returned tuple of particles is the same as $\mathbf p$ from the input state except for the interacting particles $p_j$ and $p_k$, which can be changed.
The result of the interact function $i\left( g,p_j,p_k\right)$, which is the tuple $\left(\overline p_j, \overline p_k\right)$, replaces these particles. 
\\[12pt]
In the example from Section \ref{sec:defPM:expample}:
\begin{align}
	(\overline p_1, \overline p_2) &=  i(g^1, p_1, p_2)\\
	&= i\big(g^1, (0,2), (0.49, -1)\big) \\
	&=  \big((0,-1),(0.49,2)\big),\\
	\intI ([g^1,\mathbf p^1], 1, 2) &= ( \overline p_1, \overline p_2 , p_3) \\
	&= \big((0,-1),(0.49,2), (2,1)\big). \label{eq:expl:i1-5}
\end{align}
Here, the result of the interaction of the particles $p_1$ and $p_2$ is $\overline p_1$ and $\overline p_2$. Both  $\overline p_1$ and $\overline p_2$ have the same positions as $p_1$ and $p_2$, but swapped velocities.
The function $\intI$ replaces in the particle tuple $\mathbf p$ the interacting particles $p_1$ and $p_2$ by their interaction results $\overline p_1$ and $\overline p_2$.

\subsubsection*{Explanation of the Second Interaction Sub-function $\intII$ (eqs. \ref{eq:defPM:i2}, \ref{eq:defPM:si2})}
\begin{align}
	\intII&: [G \times P^*] \times \mathbb N \rightarrow P^*   \tag{eq. \ref{eq:defPM:i2}}\\
	 \intII ([g,\mathbf p], j)&:= \mathbf p \; *_{\intI_{(g,j)}} \; u([g,\mathbf p], j) \qquad \text{with } \quad \intI_{(g,j)}(\mathbf p,k):=\intI ([g,\mathbf p], j,k) \tag{eq. \ref{eq:defPM:si2}}
\end{align}
The second interaction sub-function $\intII$ describes how the interaction of one particle with all its neighbor particles contributes to the state change of the particle method. 
It calculates the inner loop where a particle interacts with all its neighbors. 
Therefore, it uses the previous function $\intI$ and evaluates it for each neighbor given by $u$. 

Formally, it maps a state of the particle method and a  particle index to a tuple of particles.
It uses the composition operator $*$ (see Definition \ref{def:compo}) over the function $\intI_{(g,j)}$, i.e. $*_{\intI_{(g,j)}}$, to compute the loop over all neighbors of one particle.
It therefore computes the interactions of the particle $p_j$ with all its neighbors in $\mathbf p$. The neighborhood function $u([g, \mathbf p], j)$ returns the indices of these neighbors, not necessarily defined geometrically.
The result of each pairwise interaction is stored in an altered tuple of particles $\mathbf p$.
This altered tuple of particles is then used for the next pairwise interaction until $p_j$ has interacted with all its neighbors.
\\[12pt]
In the example from Section \ref{sec:defPM:expample}:
\begin{align}
        \label{eq:expl:i2-start}
	u([g^1, \mathbf p^1], 1) &= (2),\\
	 \intII ([g^1,\mathbf p^1], 1)&= \mathbf p^1 \; *_{\intI_{(g^1,1)}} \; u([g^1,\mathbf p^1], 1) \\
	 		&= \mathbf p^1 \; *_{\intI_{(g^1,1)}} \;  (2)\\
			&= \intI_{(g^1,1)}(\mathbf p^1,2) \; *_{\intI_{(g^1,1)}} \;  ()\\
			&= \intI_{(g^1,1)}(\mathbf p^1,2)\\
			&= \intI ([g^1,\mathbf p^1], 1, 2)\\
			\label{eq:expl:i2-end}
			&= \big((0,-1),(0.49,2), (2,1)\big).
\end{align}
The neighboring indices of the first particle in $\mathbf p^1$, which is $p_1$, are calculated by the neighborhood function $u([g^1,\mathbf p^1], 1)$. The result of this function is the one-element index tuple $(2)$, since only particle 2 is colliding with particle 1. 
This means that the composition operator in the function $\intII$ just needs to iterate over one index.
Composition with an empty tuple $()$ would not change anything.
Hence, the result of the function $\intI ([g^1,\mathbf p^1], 1, 2)$ as described in eq. \ref{eq:expl:i1-5}) is reduced.

\subsubsection*{Explanation of the Third Interaction Sub-function $\intIII$ (eqs. \ref{eq:defPM:i3}, \ref{eq:defPM:si3})}
\begin{align}
	  \tag{eq. \ref{eq:defPM:i3}}
	  \intIII&: [G \times P^*] \rightarrow P^*
	  \\
	  \tag{eq. \ref{eq:defPM:si3}}
	  \intIII ([g,\mathbf p])&:=\mathbf p \; *_{\intII_g} (1,..,\vert \mathbf p\vert) \qquad \text{with } \quad \intII_g (\mathbf p, j):=\intII ([g,\mathbf p], j)
\end{align}
The third interaction sub-function $\intIII$ describes how the interactions of all particles with all their respective neighbors contribute to the state change of the particle method. Therefore, it calculates both the inner and the outer loop to calculate for each particle the interaction with its neighbors, recursively using the second interaction sub-function for the inner loop.

Formally, this maps a state of the particle method to a tuple of particles.
It uses the composition operator $*$ (see Definition  \ref{def:compo}) over the second interact function $\intII_{g}$ to iterate over all indices of the particles in the tuple $\mathbf p$.
The function $\intIII$ calculates for each of these particles the interaction between these particles and all of their neighbors, hence completely computing all necessary interactions between all particles in $\mathbf p$.
\\[12pt]
In the example from Section \ref{sec:defPM:expample}:
\begin{align}
   	\label{eq:expl:i3-start}
	 \intIII ([g^1,\mathbf p^1])&=\mathbf p^1 \; *_{calculation_{g^1}} (1,2,3) \\
	 &=  \intII ([g^1,\mathbf p^1], 1) \; *_{\intII_{g^1}} (2,3) \\
	 &=  \underbrace{\big((0,-1),(0.49,2), (2,1)\big)}_{=:\hat{\mathbf p}^1} \; *_{\intII_{g^1}} (2,3) \\
	 &=  \hat{\mathbf p}^1 \; *_{\intII_{g^1}} (2,3) \\
	 &=  \intII ([g^1,\hat{\mathbf p}^1], 2) \; *_{\intII_{g^1}} (3) \\
	 &= \Big(\hat{\mathbf p}^1 \; *_{\intI_{(g^1,2)}} \; u([g^1, \hat{\mathbf p}^1], 2)\Big)  \; *_{\intII_{g^1}} (3) \\
	 &= \Big(\hat{\mathbf p}^1 \; *_{\intI_{(g^1,2)}} \; ()\Big)  \; *_{\intII_{g^1}} (3) \\
	 &= \hat{\mathbf p}^1  \; *_{\intII_{g^1}} (3) \\
	 &= \intII ([g^1,\hat{\mathbf p}^1], 3) \; *_{\intII_{g^1}} () \\
	 &= \Big(\hat{\mathbf p}^1 \; *_{\intI_{(g^1,3)}} \; u([g^1, \hat{\mathbf p}^1], 3)\Big)  \; *_{\intII_{g^1}} () \\
	 &= \Big(\hat{\mathbf p}^1 \; *_{\intI_{(g^1,3)}} \; ()\Big)  \; *_{\intII_{g^1}} () \\
	 &= \hat{\mathbf p}^1  \; *_{\intII_{g^1}} () \\
	 &= \hat{\mathbf p}^1\\
	 &= \big((0,-1),(0.49,2), (2,1)\big).
	 \label{eq:expl:i3-end}
\end{align}
The function $\intIII$ calculates the pairwise interactions of all particles with all their neighbors starting from the particle $p_1$ and its neighbors. This is already calculated in the previous step (eqs. \ref{eq:expl:i2-start} -- \ref{eq:expl:i2-end}) with the  result named $\hat{\mathbf p}^1$ for better readability.
The next particle is $p_2$. It has to interact with all of  its neighbors, calculated through $\intII ([g^1,\hat{\mathbf p}^1], 2)$. In the example, particle $p_2$ has no neighbors in $\hat{\mathbf p}^1$. Hence, there is no interaction and no change in $\hat{\mathbf p}^1$. The same is true for particle $p_3$. It has also no neighbors.
Hence, there is again no change in $\hat{\mathbf p}^1$. This makes $\hat{\mathbf p}^1$ the result particle tuple of the pairwise interactions of all particles with all their neighbors.

\subsubsection*{Explanation of the First Evolution Sub-function $\evoI$ (eqs. \ref{eq:defPM:e1}, \ref{eq:defPM:se1})}
\begin{align}
	  \tag{eq. \ref{eq:defPM:e1}}
	   \evoI&:  [G \times P^*] \times P^* \times \mathbb N \rightarrow [G \times P^*] 
	  \\
	  \tag{eq. \ref{eq:defPM:se1}}
	  \text{E.g.: } \evoI \big(  [g,\mathbf p], \mathbf q, j \big)&:= \big[ \overline g, \quad \mathbf q \circ \overline{\mathbf q}   \big]  \qquad \text{ with } \quad \left(\overline g, \overline{\mathbf q}\right) := e(g,p_j)
\end{align}
The first evolution sub-function describes how the evolution of one particle contributes to the state change of the particle method. Therefore it evolves a particle and stores the result in a new particle tuple.

Formally, it maps a state of the particle method, a particle tuple, and an index to a state of the particle method.
It handles the result of the evolution of one particle and concatenates (see Definition \ref{def:concat}) the result, a particle tuple $\overline{\mathbf q}$, to a particle tuple $\mathbf q$. Also, the function $\evoI$ handles the change this particle causes to the global variable. The calculations are done using  the evolve function $e$ of a given particle method. 
\\[12pt]
In the example from Section \ref{sec:defPM:expample}:
\begin{align}
	 \hat{\mathbf p}^1 &= \big(\hat p_1, \hat p_2, \hat p_3\big) = \big((0,-1),(0.49,2), (2,1)\big), \\
	 \left(\overline g, \overline{\mathbf q}\right)&= e(g^1,\hat p_1) \\
	 	&= \left( g^1, \; \big(( \hat x_1 + \Delta t \cdot \hat v_1, \; \hat v_1 )\big)\right)\\
		&= \left( g^1, \; \big(( 0 + 0.1 \cdot  -1, \; -1 )\big)\right)\\
		&= \left( g^1, \; \big(( -0.1, \; -1 )\big)\right), \\
	 \label{eq:expl:e1-start}
	 \evoI \big(  [g^1, \hat{\mathbf p}^1], (), 1 \big)&= \big[ \overline g, \quad () \circ \overline{\mathbf q}   \big]\\
	 	&= \big[  g^1, \quad () \circ \big(( -0.1, \; -1 )\big)  \big]\\
		&= \big[  g^1, \big(( -0.1, \; -1 )\big)  \big].
		\label{eq:expl:e1-end}
\end{align}
We use the result $\hat{\mathbf p}^1$ of the previous function $\intIII$ (eqs. \ref{eq:expl:i3-start} -- \ref{eq:expl:i3-end}) as the starting point for this example calculation, since $\evoI$ and $\intIII$ are connected. The result $\left(\overline g, \overline{\mathbf q}\right)$ of the evolve function $e(g^1,\hat p_1)$ is the initial global variable $g^1$ and the particle $\hat p_1$ with a changed  position from $\hat x_1=0$ to $\hat x_1=-0.1$.  The velocity stays the same $\hat v_1=-1$.
The function $\evoI$ uses this result to change the state $[g^1, ()]$, which consists of the initial global variable $g^1$ and an empty particle tuple $()$. The global variable $g^1$ is changed to $\overline g$ according to $e(g^1,\hat p_1)$ is $\overline g=g^1$. Hence, there is no effective change in $g^1$.
The particle tuple $\overline{\mathbf q}=\big(( -0.1, \; -1 )\big)$ is concatenated (Definition \ref{def:concat}) to the empty tuple $()$. Hence, the overall result of $\evoI$ is the state $\big[  g^1, \big(( -0.1, \; -1 )\big)  \big]$.

\subsubsection*{Explanation of the Second Evolution Sub-function $\evoII$ (eq. \ref{eq:defPM:e2}, \ref{eq:defPM:se2})}
\begin{align}
	  \tag{eq. \ref{eq:defPM:e2}}
	  \evoII&:  [G \times P^*]  \rightarrow [G \times P^*] \\
	  \tag{eq. \ref{eq:defPM:se2}}
	  \evoII \big(  [g,\mathbf p]\big) &:= \big[ g, () \big] \; *_{\evoI_{\mathbf p}} \; (1,..,\vert \mathbf p\vert)   \qquad \text{with} \quad \evoI_{\mathbf p} \big( [g,\mathbf q], j \big)  := \evoI \big( [g,\mathbf p], \mathbf q, j \big)
\end{align}
The second evolution sub-function $\evoII$ describes how the evolution of all particles contributes to the state change of the particle method. Therefore, it evolves all particles and stores the results.

Formally, it maps a state of the particle method to a state of the particle method.
It uses the composition operator $*$ (see Definition \ref{def:compo}) over the first evolution sub-function $\evoI_{\mathbf p}$ to iterate over all indices of the particles in the tuple $\mathbf p$ and compute their evolution.
The function $\evoII$ handles for each of these particles the evolution and accumulates the results in a new state starting from the state $\big[ g, () \big]$.
\\[12pt]
In the example from Section \ref{sec:defPM:expample}:
\begin{align}
\tag{eqs. \ref{eq:expl:i3-start} -- \ref{eq:expl:i3-end} }
  \hat{\mathbf p}^1 &= \big(\hat p_1, \hat p_2, \hat p_3\big) = \big((0,-1),(0.49,2), (2,1)\big),\\
 	\label{eq:expl:e2-start}
		\evoII \big(  [g^1,\hat{\mathbf p}^1]\big)&= \big[ g^1, () \big] \; *_{\evoI_{\hat{\mathbf p}^1}} \; (1,...,\vert\hat{\mathbf p}^1\vert)\\
			&= \big[ g^1, () \big] \; *_{\evoI_{\hat{\mathbf p}^1}} \; (1,2,3) \\
		 	&= \evoI \big(  [g^1, \hat{\mathbf p}^1],(), 1 \big) \; *_{\evoI_{\hat{\mathbf p}^1}} \; ( 2, 3) \\
			&= \big[  g^1, \big(( -0.1, \; -1 )\big)  \big] \; *_{\evoI_{\hat{\mathbf p}^1}} \; ( 2, 3) \\
			&= \evoI \big(  [g^1, \hat{\mathbf p}^1],\big(( -0.1, \; -1 )\big), 2 \big) \; *_{\evoI_{\hat{\mathbf p}^1}} \; (3) \\
			&= \big[  g^1, \big(( -0.1, \; -1 )\circ ( \hat x_2 + \Delta t \cdot \hat v_2, \; \hat v_2 )\big)  \big] \; *_{\evoI_{\hat{\mathbf p}^1}}  \; (3) \\
			&= \big[  g^1, \big(( -0.1, \; -1 )\circ ( 0.49 + 0.1 \cdot 2, \; 2 )\big)  \big] \; *_{\evoI_{\hat{\mathbf p}^1}} \; (3) \\
			&= \big[  g^1, \big(( -0.1, \; -1 ), ( 0.69, \; 2 )\big)  \big] \; *_{\evoI_{\hat{\mathbf p}^1}} \; (3) \\
			&= \big[  g^1, \big(( -0.1, \; -1 ), ( 0.69, \; 2 ), (2.1,\; 1)\big)  \big] .
		\label{eq:expl:e2-end}
\end{align}
We use again the result $\hat{\mathbf p}^1$ of $\intIII ([g^1,\mathbf p^1])$, which was calculated above (eqs. \ref{eq:expl:i3-start} -- \ref{eq:expl:i3-end}) for this example, knowing that $\evoII$ and $\intIII$ are connected.
Hence, it remains to calculate the evolution of the particles $\hat p_1$, $\hat p_2$, and $\hat p_3$ and add them up into the state $[g^1, ()]$. 
The evolution of $\hat p_1$, namely $\evoI \big(  [g^1, \hat{\mathbf p}^1],(), 1 \big)$, has already been calculated above (eqs. \ref{eq:expl:e1-start} -- \ref{eq:expl:e1-end}).
For the next two particles, $\hat p_2$ and $\hat p_3$, the same procedure is followed. 
After each evolution, the resulting particle tuple, with just one element, is concatenated to the previous result, such that they combine to a particle tuple of 3 particles in the end. In this example, the global variable $g^1$ remains unchanged by the function $\evoII$.

\subsubsection*{Explanation of the State Transition Function $s_\mathrm{general}$ (eqs. \ref{eq:defPM:s}, \ref{eq:defPM:ss})}
\begin{align}
	\tag{eq. \ref{eq:defPM:sgeneral}}
	s_\mathrm{general} &:  [G\times P^*] \nrightarrow [G\times P^*] 
	 \\
	  \tag{eq. \ref{eq:defPM:ss}}
	  s_\mathrm{general} \left(  [g, \mathbf p] \right)&:= 
    	\begin{cases}
    	\textit{undef} & \text{if} \qquad f\left(g\right)\\
        \big[ \mathring{e}( \overline{ g} ),\; \overline{\mathbf p}   \big]  & \text{else} 
       \end{cases} \hspace{1cm} \text{with } 
        \quad  \left[\overline g, \overline{\mathbf p}\right] := \evoII \big(  [g,\; \intIII ([g,\mathbf p])]\big) 
\end{align}
The sequential particle method state transition function brings all calculations together. Hence, it handles all the interactions and evolutions necessary to advance a particle method form one state to the next state.

Formally, the state transition function is a partial function~($\nrightarrow$) that maps a state of the particle method to a state of the particle method. If $s_\mathrm{general}$ is undefined, the particle method stops. 
$s_\mathrm{general}$ combines the functions $\intIII$, $\evoII$, $f$, and $\mathring e$ to advance the algorithm from one state to the next.
Input is the current state of the particle method $[g, \mathbf p]$ with the global variable $g$ and a tuple of particles $\mathbf p$.
The function $s_\mathrm{general}$ calculates all interactions and evolutions of the particles in $\mathbf p$ and of the global variable $g$, as defined by the interaction and evolution functions $i$ and $e$, provided the stopping condition $f(g)=\bot$ (false). Otherwise, there is no next state of the particle method.
This is the reason why $s_\mathrm{general}$ is a partial function ($\nrightarrow$).
If $f(g)$ is false, $s_\mathrm{general}$ returns the new state, which consists of the result of the evolve function of the global variable $\mathring e(\overline g)$ and of the new tuple of particles $\overline{\mathbf p}$.
The state $[\overline g, \overline{\mathbf p}]$ is the result of the calculation of all interactions and all evolutions of the particles in $\mathbf p$, as calculated by $\evoII \big(  [g,\; \intIII ([g,\mathbf p])]\big)$.
\\[12pt]
In the example from Section \ref{sec:defPM:expample}:
\begin{align}
	\tag{eq. \ref{eq:expPMI:g}}
	g^1 &= (d, t, \Delta t, T) = (0.5, 0, 0.1, 10) ,
	\\
	\tag{eq. \ref{eq:expPMI:p}}
	 \mathbf p^1 & = \big( (0,2), (0.49, -1), (2,1) \big), 
	 \\
	\tag{eq. \ref{eq:expl:e2-start} -- \ref{eq:expl:e2-end} }
	\evoII \big(  [g^1,\; \intIII ([g^1,\mathbf p^1])]\big)&= \big[  g^1,  \underbrace{\big(( -0.1, \; -1 ), ( 0.69, \; 2 ), (2.1,\; 1)\big)}_{=:\mathbf p^2}  \big], 
\end{align}

\begin{align}
	f(g^1)&= (t \geq T)\\
	&= (0 \geq 10)\\
	&=\bot ,
	\\
	\label{eq:expl:s-start}
	s_\mathrm{general} \left(  [g^1, \mathbf p^1] \right)&:= 
    	\begin{cases}
    	\textit{undef} & \text{if} \qquad f\left(g^1\right)\\
        \big[ \mathring{e}( \overline{ g} ),\; \overline{\mathbf p}   \big]  & \text{else} 
       \end{cases} \quad \text{ with} 
        \quad  \left[\overline g^1, \overline{\mathbf p}\right] := \evoII \big(  [g,\; \intIII ([g^1,\mathbf p^1])]\big) \\
        \\
 	s_\mathrm{general} \left(  [g^1, \mathbf p^1] \right)
	&= 
	\begin{cases}
		\textit{undef} & \text{if} \qquad f\left(g^1\right)\\
    		\big[ \mathring{e}( \overline{ g} ),\; \overline{\mathbf p}   \big] \qquad \text{with } 
    		\quad \left[\overline g, \overline{\mathbf p}\right] := \evoII \big(  [g^1,\; \intIII ([g^1,\mathbf p^1])]\big) & \text{else} 
   	\end{cases}\\
	&= 
	\begin{cases}
		\textit{undef} & \text{if} \qquad \bot\\
    		\big[ \mathring{e}( \overline{ g} ),\; \overline{\mathbf p}   \big] \qquad \text{with } 
		\quad \left[\overline g, \overline{\mathbf p}\right] =[g^1, \mathbf p^2] & \text{else} 
   	\end{cases}\\
	&= \big[ \mathring{e}(  g^1 ),\; \mathbf p^2  \big] \\
	&= \big[ (d, t+\Delta t, \Delta t, T),\; \mathbf p^2  \big] \\
	&= \big[ (0.5, 0+0.1, 0.1, 10),\; \mathbf p^2  \big] \\
	&= \big[ \underbrace{(0.5, 0.1, 0.1, 10)}_{=: g^2},\; \mathbf p^2  \big] \\
	&= \big[ g^2,\; \mathbf p^2  \big] .
	\label{eq:expl:s-end}
\end{align}
The starting state $[g^1, \mathbf p^1]$ in this example is given as a reminder, as well as the second evolve sub-function $\evoII \big(  [g^1,\; \intIII ([g^1,\mathbf p^1])]\big)$. We define the particle tuple of the result of $\evoII$ as $\mathbf p^2$. This indicates that this particle tuple is already part of the next state $[g^2, \mathbf p^2]$. This is true because there will be no further change to it.
The stopping condition $f$ determines whether the particle method stops, i.e., if $f(g)=\top$. Then, there is no next state. If $f(g)=\bot$, as in this example, then there is a next state. Since $\evoII \big(  [g^1,\; \intIII ([g^1,\mathbf p^1])]\big)=[g^1, \mathbf p^2]$, only the evolution of the global variable $\mathring e$ remains. 
$\mathring{e}(  g^1 )$ advances the current time $t$ by one time step size $\Delta t$. 
Hence, the current time increases from $t=0$ to $t=0.1$. The rest of the properties of the global variable remain unchanged in this example. We call the new global variable  $g^2$ because there is no further change in this step.

\subsubsection*{Explanation of the Next State $[ g^{t+1}, \mathbf p^{t+1}]$ (eq. \ref{eq:defPM:tpp})}
\begin{align}
	   [ g^{t+1}, \mathbf p^{t+1}] &:= s_{general}\left(  [g^{t}, \mathbf p^{t}]\right)  \tag{ eq. \ref{eq:defPM:tpp}}
\end{align}
The state transition function applied to the current state $[g^t, \mathbf p^t]$ advances the particle method to its next state $[g^{t+1}, \mathbf p^{t+1}]$. This is iterated until the stopping state is reached (see Definition \ref{def:PMS}, eq. \ref{eq:defPMS:lastS}). 
\\[12pt]
In the example from Section \ref{sec:defPM:expample}:
\begin{align}
	\tag{eqs. \ref{eq:expl:s-start}--\ref{eq:expl:s-end}}
	\big[ g^2,\; \mathbf p^2  \big]  
		&= s_{general} \left(  [g^1, \mathbf p^1] \right)\\
		&= s_{general} \left(  [(0.5, 0, 0.1, 10),\; \big( (0,2), (0.49, -1), (2,1) \big) ] \right) \notag\\
		&=  \left[(0.5, 0.1, 0.1, 10),\: \big(( -0.1, \; -1 ), ( 0.69, \; 2 ), (2.1,\; 1)\big)\right]. \hspace{4cm} \notag
\end{align}
In the example, all calculations are done using the state transition function $s_\mathrm{general}$ of the particle method (eqs. \ref{eq:expl:s-start} -- \ref{eq:expl:s-end}). This iterates until the state becomes undefined when the stopping time has been reached, as defined by the stopping condition $f$.

\section{Examples} \label{sec:example}
We show how classic particle methods can be formalized 
using our definition by considering the examples of Particle Strength Exchange (PSE) as a continuous particle method and Molecular Dynamics (MD) as a discrete particle method. We further show how also other algorithms, not generally recognized as particle methods, can be cast in terms of our definition. As examples, we consider triangulation refinement and Gaussian elimination. 

\subsection{Particle Strength Exchange}
Particle Strength Exchange (PSE) is a classic particle method to numerically solve partial differential equations in time and space \cite{Degond:1989, Degond:1989a, Eldredge:2002, Schrader:2010}. It provides a general framework for numerically approximating differential operators over sets of irregularly placed collocation points, called particles.
Here, we consider the example of using PSE to numerically solve the isotropic, homogeneous, and normal diffusion equation in one dimension:
\begin{equation}
\frac{\partial w(x,t)}{\partial t} = D   \frac{\partial^2 w(x,t)}{\partial x^2} \label{eq:diff1d}
\end{equation}
for the continuous and sufficiently smooth function $w(x,t) : \mathbb{R}^2\to\mathbb{R}$. 
For time integration, we again use the explicit Euler method, whereas space is discretized using PSE on  equidistant points with spacing $h$.
This approximates the second derivative in space  $\frac{\partial^2 w}{\partial x^2}$ at location $x_j$ using the surrounding particles at positions $x_k$ as \cite{Degond:1989}:
\begin{equation}
 \frac{\partial^2 w}{\partial x^2}(x_j) \approx \frac{h}{\epsilon^2} \sum^N_{k=1} \left(w(x_k)-w(x_j)\right) \eta_{\epsilon}(x_k-x_j).
\end{equation}

Using PSE theory, we determine the operator kernel $\eta_{\epsilon}$ such as to yield an approximation error that converges with the square of the kernel width $\epsilon$:
\begin{equation}
\eta_{\epsilon}(x):=\frac{1}{2 \epsilon \sqrt{\pi}}\exp \left( \frac{-x^2}{4 \epsilon^2}\right).
\end{equation}
The support of the kernel is $[-\infty, \infty]$. However, the exponential quickly drops below the machine precision of a digital computer, so it is custom to introduce a cut-off radius $r_c$ to limit particle interactions to non-trivial computations.
The approximation of the second derivative in space then is:
\begin{equation}
 \frac{\partial^2 w}{\partial x^2}(x_j) \approx \frac{h}{2 \epsilon^3 \sqrt{\pi}} \sum_{x_k : \: 0<\vert x_k-x_j\vert \leq r_c} \left(w(x_k)-w(x_j)\right) \exp \left( \frac{-(x_k-x_j)^2}{4 \epsilon^2} \right). \label{exmp:diffusion:PSEaprox}
\end{equation}

Time is discretized using the explicit Euler method as introduced in Section \ref{sec:defPM:expample}. 
The explicit Euler method allows the approximation of the continuous time derivative  $\frac{\partial w}{\partial t}$ at discrete points in time $t_n$, $n\in \mathbb N$ with time step size $\Delta t:= t_{n+1}-t_n$:
\begin{equation}
 \frac{\partial w}{\partial t}(t_n) \approx \frac{w(t_{n+1})-w(t_{n})}{\Delta t}
\end{equation}
Hence, the above differential equation is discretized as:
\begin{align}
w(x_j, t_{n+1}) \approx& w(x_j, t_n)+ D \Delta t \frac{\partial^2 w}{\partial x^2}(x_j)\\
 \approx& w(x_j, t_n)+ \frac{D h \Delta t }{2 \epsilon^3 \sqrt{\pi}} \sum_{x_k : \: 0<\vert x_k-x_j\vert \leq r_c} \left(w(x_k, t_n)-w(x_j, t_n)\right) \exp \left( \frac{-(x_k-x_j)^2}{4 \epsilon^2} \right). \label{exmp:diffusion:euler}
\end{align}

In order to numerically solve eq. \ref{eq:diff1d}, this expression is evaluated over the particles at locations $x_j$ with property $w_j$ at time points $t_n$.
For simplicity, we consider a free-space simulation without boundary conditions.
Hence, we assume that an instance of this particle method has enough particles with no or low concentration $w_j$ around the region of interest in the initial tuple of particles. We further assume that the particles are regularly spaced with inter-particle spacing $h$ such that $\frac{h}{\epsilon} \leq 1$ (this is a theoretical requirement in PSE known as the ``overlap condition''. Without it, the numerical method is not consistent). This defines the particle method:
\begin{align}
 	 p:=&\left(x, w, \Delta w\right) \quad \text{ for }p \in P := \mathbb{R} \times \mathbb{R} \times \mathbb{R} ,	\\
	 g:=&\left(D, h, \epsilon, r_c, \Delta t, T , t \right) \quad \text{ for } g \in  \mathbb{R}^7 , \\
	u \left([g, \mathbf p], j \right):= &\left(k \in \mathbb N: \; p_k, p_j \in \mathbf p \;\land\; 0 < \vert x_k-x_j\vert \leq r_c \right) , \\
	 f(g) :=& \left(t \geq T\right), \\
	 i \left(g, p_j,p_k \right):=& \left(
	 \left(
	 \begin{matrix}
    	&   x_j\\
    	& w_j\\
    	&  \Delta w_j + (w_k-w_j) \exp \left( \frac{-(x_k-x_j)^2}{4 \epsilon^2}\right)
	 \end{matrix}
	 \right)^\mathbf T, \; p_k\right), \\
	 e \left( g, p_j \right):= &\left( g,
	 \left( \left(
	 \begin{matrix}
    	&\vspace{2mm}  x_j\\
        &\vspace{2mm}  w_j+ \Delta t \frac{D h}{2 \epsilon^3 \sqrt{\pi}} \Delta w_j\\
        & 0
	 \end{matrix}
	 \right)^\mathbf T \right)\;
	 \right),\\
	\mathring{e} \left( g \right):=&\left(D, h, \epsilon, r_c, \Delta t, T , t+\Delta t \right).
\end{align}
Each particle $p$ represents a collocation point of the numerical scheme.
It is a collection of three properties, each of which a real number: the position $x$, the concentration $w$, and the accumulator variable $\Delta w$ that collects the concentration change through the interactions $i$. An accumulator variable is required in this case to render the result of the computation independent of the indexing order of the particles. 

The global variable $g$ is a collection of seven real-valued properties that are accessible throughout the whole calculation: the diffusion constant $D$, the spacing between particles $h$, the kernel width $\epsilon$, the cut-off radius $r_c$, the time step size $\Delta t$, the end time of the simulation $T$, and the current time $t$.

The neighborhood function $u$ returns the surrounding particles no further away than the cut-off radius $r_c$ and different from the query particle itself.
The stopping condition $f$ is true ($\top$) if the current time $t$ reaches or exceeds the end time $T$.
Then the simulation halts.

The interact function $i$ evaluates the sum in the PSE approximation (eq. \ref{exmp:diffusion:PSEaprox}).
Each particle $p_j$ accumulates its concentration change in  $\Delta w_j$ during the interactions with the other particles.
In the present example, we choose an asymmetric interact function $i$, just changing particle $p_j$.
The neighborhood function $u$ accounts for this. This is not necessary, and symmetric formulations of PSE are also possible.

The evolve function $e$ uses the accumulated change $\Delta w_j$ to update the concentration $w_j$ of particle $p_j$ using the explicit Euler method (eq. \ref{exmp:diffusion:euler}).
For that, it also uses $D$, $h$, $\epsilon$, and $\Delta t$ from the global variable $g$.
In addition, the evolve function $e$ resets the accumulator $\Delta w_j$ to $0$.
In this example, the evolve function does not change the global variable $g$. That is exclusively done in $\mathring e$, which updates the current time $t$ by adding the time step size $\Delta t$ into it.

\begin{figure}[t]
\centering
\includegraphics[scale=0.5]{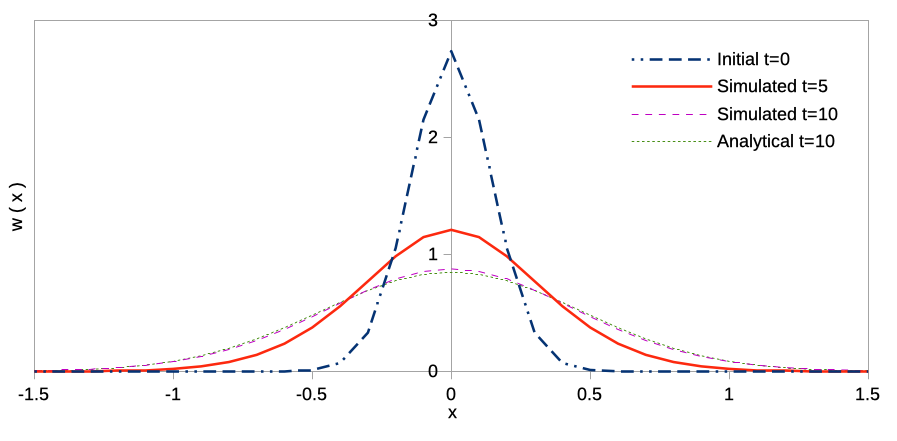}
\caption{Particle Strength Exchange simulation of diffusion in one dimension at the times $t=0 , 5 , 10$ compared with the analytical solution at $t=10$.}
\label{fig:diffusion} 
\end{figure}

In order to define a certain instance of this particle method, we need to fix the parameters and the initial condition. 
For example, we could fix the initial concentration by discretizing the function $w(x)= \frac{1}{\sqrt{4 \pi D }}  \exp\left(\frac{-x^2}{4 D }\right) $ in the domain $x\in [-1.5 , 1.5]$ and set the parameters as:
\begin{align*}
g^1:=&\left(D, h, \epsilon, r_c, \Delta t, T , t \right) &&\text{initial global variable}\\
D&:=0.01 &&\text{diffusion constant}\\
h&:=0.1 &&\text{spacing between particles}\\
\epsilon&:=h &&\text{PSE kernel width}\\
r_c&:= 4 h &&\text{interaction cut-off radius}\\
\Delta t&:= 0.1 &&\text{time step size}\\
T &:=10 &&\text{end time of the simulation}\\
t&:=0 &&\text{current time}\\
\\
\mathbf p^1:=& (p_1,...,p_{31}) &&\text{initial particle tuple}\\
p_j &:=  (x_j, w_j, \Delta w_j) &&\text{j-th particle}\\
x_j &:=(j-1) h-1.5 &&\text{equidistant particle positions}\\
w_j &:= \frac{1}{\sqrt{4 \pi D }}  \exp\left(\frac{-x_j^2}{4 D }\right) &&\text{concentrations} \\
 \Delta w_j &:=\; 0 &&\text{accumulator}
\end{align*}
The result of executing this instance is visualized in Fig. \ref{fig:diffusion} for times $t=0,5,10$ and compared with the exact, analytical solution of this one-dimensional diffusion problem at the final time $t=T=10$.

\subsection{Lennart-Jones Molecular Dynamics}
Particle methods cannot only be used to discretize continuous models, such as the PDE from the previous example, but also to simulate discrete models. A famous example is molecular dynamics, where the Newtonian mechanics of a collection of discrete atoms or molecules is simulated by evaluating their interaction forces. 
A classic molecular dynamics simulation is that of a so-called Lennart-Jones gas, a dilute collection of  electrically neutral, inert, and mono-atomic molecules that interact with each other according to the Lennard-Jones potential~\cite{Lennard-Jones:1931}
\begin{equation}
U(r)=4 \epsilon \left( \left(\frac{\sigma}{r}\right)^{12}- \left(\frac{\sigma}{r}\right)^{6}\right).
\end{equation}
This potential defines the energy of the interaction as a function of the distance $r$ between two atoms. It is a good approximation of how noble gases behave in function of two parameters: $\sigma$ that controls the equilibrium distance between atoms and hence the taget density of the gas, and $\epsilon$ that defines the strength of the interactions by setting the energy scale of the potential. 

From this interaction potential, the force between any two atoms a distance $r$ apart can be computed as $F = - \frac{\mathrm{d}U(r)}{\mathrm{d}r}$. From the forces summed up over all other atoms, the acceleration of each atom is computed from Newton's law $F=ma$, where $m$ is the mass of the atom. The acceleration is then used to update the velocity and position of the atoms using a symplectic (i.e., energy conserving) time-stepping method, such as the velocity Verlet scheme~\cite{Verlet:1967,Swope:1982}.

For simplicity, we choose $\epsilon=\sigma=m=1$.
The acceleration between two atoms $j$ and $k$ a distance $r_{jk}:=x_k-x_j$ apart then is:
\begin{equation}
a(r_{jk})=\frac{24}{r_{jk}^{7}}- \frac{48}{r_{jk}^{13}}.
\end{equation}
This acceleration is then used in the velocity-Verlet method~\cite{Verlet:1967} to update the velocity $v$ and position $x$ of an atom over a time step $\Delta t$, from $t_n$ to $t_{n+1}$, as:
\begin{align}
&\text{compute } a(t_{n}) \text{ from } x(t_{n}),\\
&v(t_{n})= v(t_{n-\frac{1}{2}}) + \frac{\Delta t}{2} \; a(t_n),\\
&v(t_{n+\frac{1}{2}})= v(t_{n}) + \frac{\Delta t}{2} \; a(t_n),\\
&x(t_{n+1} )= x(t_n) + \Delta t \; v(t_{n+\frac{1}{2}}).\label{eq:example:MD:x}
\end{align}
We choose to have periodic boundary conditions at the boundaries of the simulation domain $x \in [0,D)$. These  boundary conditions are imposed by a directed distance function
\begin{align}
    d_D(x,y):=
    \begin{cases}
       y-x - D \qquad&\text{if}\quad y-x > \;\; \frac{1}{2} D, \\
       y-x + D       &\text{if}\quad y-x \leq  -\frac{1}{2} D, \\
       y-x &\text{else}
    \end{cases}
\end{align}
and through the modulo operator
\begin{align}
a\hspace{-3mm} \mod \; b := r  \quad\Longleftrightarrow\quad a=b \cdot c + r \qquad\text{with}\quad a,b \in \mathbb R,\; c\in \mathbb Z,\; r\in \left[0, |b|\right)  ,
\end{align}
that we use to adapt equation \ref{eq:example:MD:x} to
\begin{align}
x(t_{n+1} )= \left(x(t_n) + \Delta t \; v(t_{n+\frac{1}{2}})\right)\hspace{-3mm}\mod{D}.
\end{align}

In a particle method simulation of Lennard-Jones molecular dynamics, each atom is represented as a distinct particle. They interact pairwise to compute the resultant sum of all  accelerations of each atom and then evolve their position and velocity using velocity-Verlet time stepping. For simplicity, we again consider a one-dimensional domain without boundaries.
This defines the following particle method in our framework:
\begin{align}
 	 p:=&\left(x, v, a  \right) \quad \text{ for }p \in P := \mathbb{R}^3,	\\
	 g:=&\left( r_c, D, \Delta t, T , t \right) \quad \text{ for } g \in  \mathbb{Q}^5,\\
	u \left([g, \mathbf p], j \right):= &(k \in \mathbb N: \; p_k, p_j \in \mathbf p \;\land\; 0 <  d_D(x_j, x_k) \leq r_c ),\\
	  f(g) :=& \left(t \geq T\right),\\
	 i \left(g, p_j, p_k \right):=& \left( \left( 
	 \begin{matrix}
    	& x_j\\
    	& v_j\\
    	&\vspace{2mm} a_j+\frac{24}{d_D(x_j, x_k)^7}-\frac{48}{d_D(x_j, x_k)^{13}}
	\end{matrix}
	\right)^\mathbf T,
	\left(
	\begin{matrix}
    	& x_k\\
    	& v_k\\
    	&\vspace{2mm} a_k+\frac{24}{d_D(x_k, x_j)^7}-\frac{48}{d_D(x_k, x_j)^{13}}
	\end{matrix}
	\right)^\mathbf T\; \right),
	\\
	 e \left(g, p_j \right):=& \left( g,
	 \left(\left(
	 \begin{matrix}
    	 & \left( x_j + \Delta t \; \left(v_j +  \Delta t \cdot a_j   \right) \right)\hspace{-3mm} \mod \;D \quad  \\
    	 & v_j +  \Delta t \cdot a_j  \\
    	 & 0
	\end{matrix}
	\right)^\mathbf T \;\right)\right),\\
	\mathring{e} \left( g \right):=&\left( r_c, D, \Delta t, T , t+\Delta t \right).
\end{align}

\begin{figure}[t]
\centering
\includegraphics[width=\textwidth]{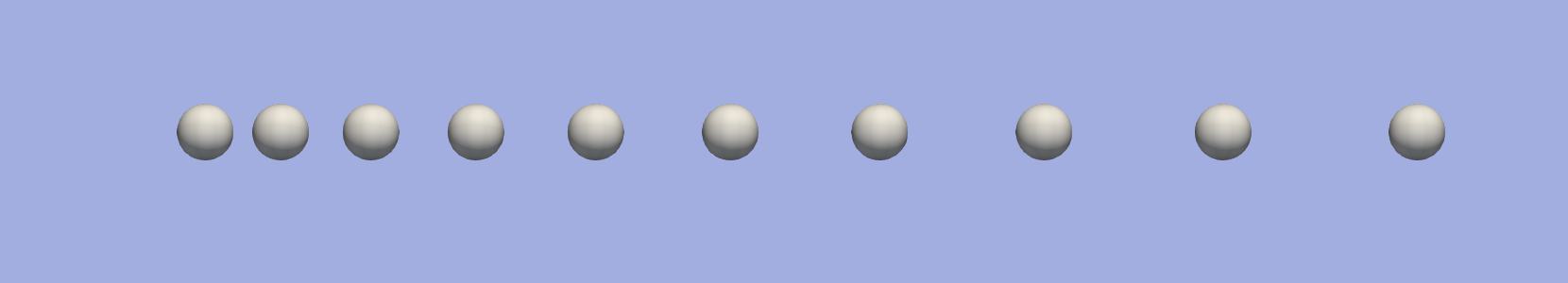}
\includegraphics[width=\textwidth]{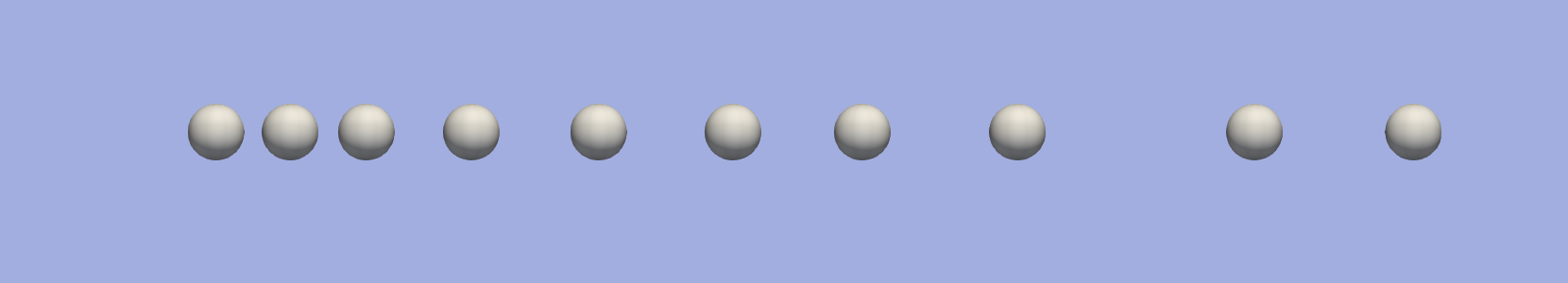}
\caption{Visualization of a molecular dynamics simulation of 10 Lennart-Jones atoms in one dimension at the times $t=0$ (top) and $t=T=10$ (bottom).}
\label{fig:MD} 
\end{figure}

A particle $p$ is in this example the collection of properties of one atom, namely its position $x$, velocity $v$, acceleration $a$, and previous acceleration $a^{old}$, as required for the velocity-Verlet method.
All properties are real numbers.
The global variable $g$ is a collection of rational numbers: the cut-off radius $r_c$ for the interactions, the time-step size $\Delta t$, the end time of the simulation $T$, and the current time $t$.

The neighborhood function $u$ returns the surrounding particles which are no further away than the cut-off radius $r_c$, are not the particle itself, and are to the right  of the particle (absence of absolute value in the distance computation). This hence defines an asymmetric neighborhood, which is used for symmetric interactions.
The stopping condition $f$ is true ($\top$) if the current time $t$ reaches or exceeds the end time $T$.

The interact function $i$ sums all forces acting on a particle.
Because the neighborhood is asymmetric, the interactions are symmetric, hence changing both involved particles.
In each pairwise interaction of two particles $p_j$ and $p_k$, these particles add the contribution of the Lennart-Jones acceleration into their current acceleration. 
The evolve function $e$ computes for a particle $p_j$ the new position $x_j$ and the new velocity $v_j$ from the current acceleration $a_j$ (computed in $i$) and the previous acceleration $a_j^{old}$ (stored).
For this, it also uses the time step size $\Delta t$ from the global variable $g$.
In the velocity-Verlet method, the position $x$ is calculated before the acceleration $a$. Since the acceleration $a$ is calculated in $i$, the position $x$ needs to be calculated afterward. Hence, $x$ for the next time step is calculated in $e$ using $a(t_{n+1}$ and $v(t_{n+1}$.
The evolve function also overwrites $a_j^{old}$ with $a_j$ and resets $a_j$ to $0$ for the next time step.
In this example, it does not change the global variable $g$.
That is only done in $\mathring e$, which advances the current time $t$ by adding the time step size $\Delta t$.

\begin{figure}[t]
\centering
\includegraphics[width=\textwidth]{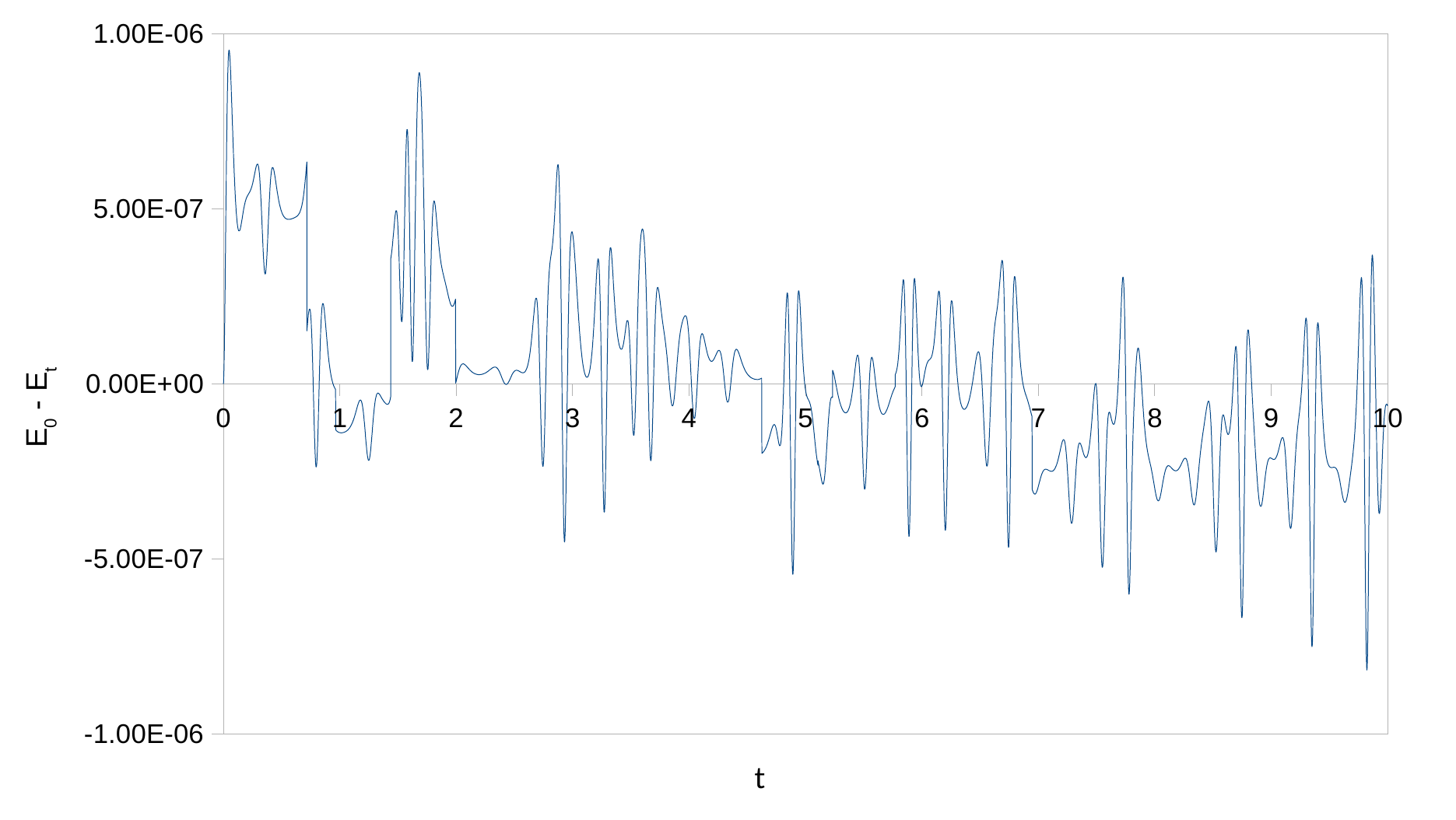}
\caption{Deviation of the total energy of the molecular dynamics simulation from the total initial energy at $t=0$.}
\label{fig:MDEnergy} 
\end{figure}

An instance is again defined by fixing the parameters and initial state of the particle method. In this example, we choose to initially place 10 particles with a linearly increasing spacing between them and initialize all other properties to 0, hence:
\begin{align*}
g^1:=&\left( r_c, D, \Delta t, T , t \right) &&\text{initial global variable}\\
r_c&:= 3 &&\text{cut-off radius}\\
D&:= 19 &&\text{domain size}\\
\Delta t&:= 0.0001 &&\text{time step size}\\
T &:=10 &&\text{end time of the simulation}\\
t&:=0 &&\text{current time}\\
\\
\mathbf p^1&:= (p_1,...,p_{10}) &&\text{initial particle tuple}\\
p_j := & (x_j, v_j, a_j) &&\text{j-th particle}\\
x_j:=& j  (0.9+0.11 j) &&\text{initial particle positions}\\
v_j:= &0 &&\text{initial particle velocities}\\
a_j:=& 0 &&\text{initial particle accelerations}
\end{align*}
The result of executing this instance is shown in Fig. \ref{fig:MD} at initial and final time. As a validation, we show in Fig. \ref{fig:MDEnergy} that the total (i.e., kinetic plus potential) energy of the system is behaving as expected for a symplectic time integrator like the velocity-Verlet scheme used here. Due to numerical round-off errors, the global energy difference fluctuates around zero with errors comparable to the square of machine epsilon for double-precision floating-point arithmetics, as expected from the second-order time integration method used.

\subsection{Triangulation refinement}
After having seen two classic particle methods, one for simulating a continuous model and one for simulating a discrete model, we also show two examples of how algorithms not generally recognized as particle methods can be formulated in our definition. This is not to imply that one should actually implement them like this in practice, but to show the generality of our definition and the value of having a formal framework. 

\begin{figure}[t]
\centering
\includegraphics[scale=0.5]{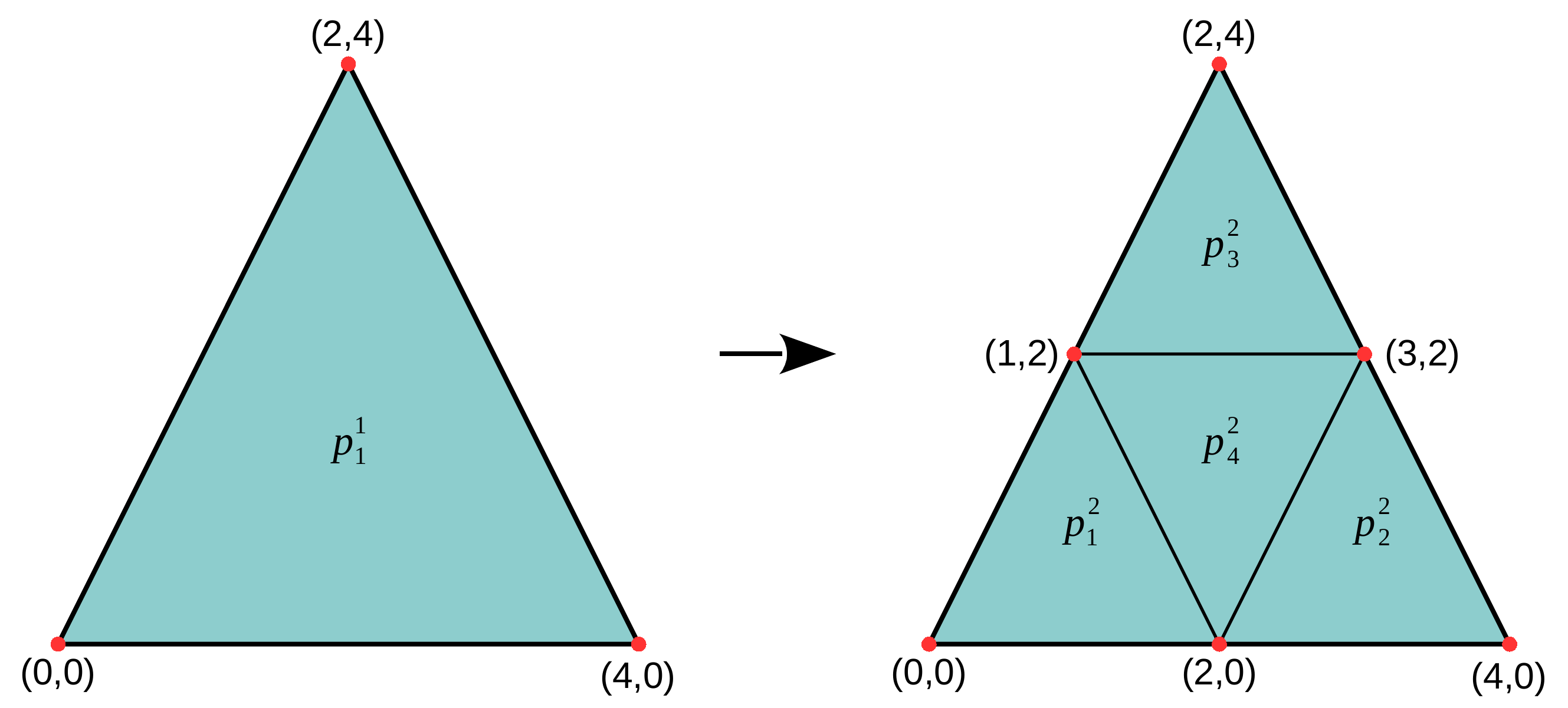}
\caption{Visualization of the \emph{refinement} of a single triangle $\mathbf p^1$ to four triangles by inserting three new vertices.}
\label{fig:triangulation} 
\end{figure}

The first example considers an algorithm for triangulation refinement as often used in computer graphics~\cite{Despreaux:2018}. It refines a triangulation by replacing each triangle with four smaller triangles.
It does so by first creating three new vertices for each existing triangle, one at the midpoint of each edge of the triangle. 
These new vertices are then connected with three new edges.
Together with the already existing vertices and edges, this creates the refined triangulation. The process is illustrated in Fig.~\ref{fig:triangulation}.

There are multiple ways of formulating this process as a particle method. An obvious choice might be to represent each vertex of the triangular mesh by a particle, and to store the incoming or outgoing edges as (integer-valued) properties of the particles. Another choice, which we follow here for the sake of illustration, is to represent each triangle by a particle. 
Each particle (triangle) then stores a unique index, a vector of three vertices, the indices of its face-connected neighbors, and a ``reverse index'' storing the information which neighbor (0th, 1st, or 2nd) it is from the perspective of its neighbor particles.

Through the interact function $i$, these neighbor information are exchanged.
In the evolve function $e$, they are then used to calculate the neighbor indices of the new, smaller particles (triangles).
The old particles (triangles) are deleted by the evolve function $e$ and four new particles (triangles) are created. We choose this representation because it illustrates a case where the evolve function creates and destroys particles. Assuming that everything happens in a two-dimensional space, the resulting particle method reads:

\begin{align}
p:=&\left(\iota, \left(
\begin{array}{c}
{}^{0}v\\
{}^{1}v\\
{}^{2}v\\
\end{array}
 \right),  
 \left(
\begin{array}{c}
{}^{0}\beta\\
{}^{1}\beta\\
{}^{2}\beta\\
\end{array}
 \right),
 \left(
\begin{array}{c}
{}^{0}\gamma\\
{}^{1}\gamma\\
{}^{2}\gamma\\
\end{array}
 \right) \right) \quad \text{ for } p \in P := \mathbb{N}_0 \times \left({\mathbb{R}^2}\right)^3 \times (\mathbb{N}_0 \cup \{-1\} ) ^3  \times \{0,1,2\}^3 ,
 \\[12pt]
g:=& \left( T, t\right) \quad \text{ for } g \in G := \mathbb{N}_0 \times  \mathbb{N}_0, 
\\[12pt]
u \left([g, \mathbf p], j \right):=& \left( {}^{0}\beta_j, {}^{1}\beta_j, {}^{2}\beta_j  \right),
\\[12pt]
 f(g) :=& \left(t \geq T\right),
\end{align}  

\begin{align}
&i(g, p_j, p_k):=\\
&\left( \left(\iota_j, \left(
\begin{array}{c}
{}^{0}v_j\\
{}^{1}v_j\\
{}^{2}v_J\\
\end{array}
 \right),  
 \left(
\begin{array}{c}
{}^{0}\beta_j\\
{}^{1}\beta_j\\
{}^{2}\beta_j\\
\end{array}
 \right),
 \begin{cases}
      \left(\begin{array}{c}
	r\\
	{}^{1}\gamma_j\\
	{}^{2}\gamma_j\\
     \end{array} \right)
      \;\text{ with }\; \iota_j = {}^{r}\beta_k &\quad \text{ if }\; \iota_k= {}^{0}\beta_j 
 \\
       \left(\begin{array}{c}
	{}^{0}\gamma_j\\
	r\\
	{}^{2}\gamma_j\\
     \end{array} \right)
      \;\text{ with }\; \iota_j = {}^{r}\beta_k &\quad \text{ if }\; \iota_k= {}^{1}\beta_j 
 \\
      \left(\begin{array}{c}
	{}^{0}\gamma_j\\
	{}^{1}\gamma_j\\
	r
     \end{array} \right)
      \;\text{ with }\; \iota_j = {}^{r}\beta_k &\quad \text{ if }\; \iota_k= {}^{2}\beta_j 
 \end{cases}
  \right), p_k \right),\notag
\\[6pt]
   &e(g,p_j):=\\
   &\left( g,
   \left(
   \begin{matrix}
         &\vspace{3mm} \left( 4 \; \iota_j, \left(
        \begin{array}{c}
            {}^{0}v_j\vspace{3mm}\\
            \frac{{}^{0}v_j+{}^{1}v_j}{2}\vspace{3mm}\\
            \frac{{}^{0}v_j+{}^{2}v_j}{2}\\
        \end{array}
         \right),  
         \left(
        \begin{array}{l}
        \begin{cases}
            4 \; {}^{0}\beta_j+\,({}^{0}\gamma+1) \hspace{-3mm} \mod 3 & \text{if} \quad  {}^{0}\beta_j\neq-1\\
            -1& \text{else}
           \end{cases}\vspace{1mm}\\
        \, 4 \iota_j +3\vspace{1mm}\\
        \begin{cases}
            4\; {}^{2}\beta_j+{}^{2}\gamma & \text{if} \quad \; {}^{2}\beta_j\neq-1\\
            -1& \text{else}
         \end{cases}\\
        \end{array}
         \right),
         \left(
        \begin{array}{c}
        0\\
        0\\
        0\\
        \end{array}
         \right) \right)
        \\
         & \vspace{3mm} \left( 4 \; \iota_j + 1, \left(
        \begin{array}{c}
        {}^{1}v_j\vspace{3mm}\\
        \frac{{}^{1}v_j+{}^{2}v_j}{2}\vspace{3mm}\\
        \frac{{}^{0}v_j+{}^{1}v_j}{2}\\
        \end{array}
         \right),  
         \left(
        \begin{array}{l}
        \begin{cases}
            4 \; {}^{1}\beta_j+\,({}^{1}\gamma+1) \hspace{-3mm} \mod 3 & \text{if} \quad  {}^{1}\beta_j\neq-1\\
            -1& \text{else}
           \end{cases}\vspace{1mm}\\
        \, 4 \iota_j +3\vspace{1mm}\\
        \begin{cases}
            4\; {}^{0}\beta_j+{}^{0}\gamma & \text{if} \quad \; {}^{0}\beta_j\neq-1\\
            -1& \text{else}
         \end{cases}\\
        \end{array}
         \right),
         \left(
        \begin{array}{c}
        0\\
        0\\
        0\\
        \end{array}
         \right) \right)
        \\
        & \vspace{3mm}\left( 4 \; \iota_j+2, \left(
        \begin{array}{c}
        {}^{2}v_j\vspace{3mm}\\
        \frac{{}^{0}v_j+{}^{2}v_j}{2}\vspace{3mm}\\
        \frac{{}^{1}v_j+{}^{2}v_j}{2}\\
        \end{array}
         \right),  
         \left(
        \begin{array}{l}
        \begin{cases}
            4 \; {}^{2}\beta_j+\,({}^{2}\gamma+1) \hspace{-3mm} \mod 3 & \text{if} \quad  {}^{2}\beta_j\neq-1\\
            -1& \text{else}
           \end{cases}\vspace{1mm}\\
        \, 4 \iota_j +3\vspace{1mm}\\
        \begin{cases}
            4\; {}^{1}\beta_j+{}^{1}\gamma & \text{if} \quad \; {}^{1}\beta_j\neq-1\\
            -1& \text{else}
         \end{cases}\\
        \end{array}
         \right),
         \left(
        \begin{array}{c}
        0\\
        0\\
        0\\
        \end{array}
         \right) \right)
        \\
         &  \left(4 \; \iota_j+3, \left(
        \begin{array}{c}
        \frac{{}^{0}v_j+{}^{1}v_j}{2}\vspace{2mm}\\
        \frac{{}^{1}v_j+{}^{2}v_j}{2}\vspace{2mm}\\
        \frac{{}^{0}v_j+{}^{2}v_j}{2}\\
        \end{array}
         \right),  
         \left(
        \begin{array}{c}
        4 \; \iota_j+1\vspace{1mm}\\
        4 \; \iota_j+2\vspace{1mm}\\
        4 \; \iota_j\\
        \end{array}
         \right),
         \left(
        \begin{array}{c}
        0\\
        0\\
        0\\
        \end{array}
         \right) \right) 
 \end{matrix}
 \right)^\mathbf T \;\right),\notag
 \\[6pt]
 &\mathring e(g):= \left( T,  t+1\right).
\end{align}

In this example, a particle $p$ is the collection of properties of one triangle: the index $\iota$, the vertices $\left({}^{0}v, {}^{1}v, {}^{2}v \right)^T$, the indices of the neighbor particles (triangles) $\left({}^{0}\beta, {}^{1}\beta, {}^{2}\beta \right)^T$, and the storage for the information which neighbor (0th, 1st, or 2nd) the particle is from the perspective of its neighbor particles $\left({}^{0}\gamma, {}^{1}\gamma, {}^{2}\gamma \right)^T$.
The index is a natural number. 
Each vertex is a real 2-vector storing the position of the vertex in a 2D plane. 
The neighbor indices are natural numbers as well, but including $-1$ to indicate if there is no neighbor.
The storage order is $0$, $1$, $2$.
The global variable $g$ is a collection of two natural numbers including $0$: the number of refinement steps $T$ to be made and the current refinement step $t$.

The neighborhood function $u$ simply returns the indices of the neighbor particles (triangles) $\left({}^{0}\beta, {}^{1}\beta, {}^{2}\beta \right)^T$. 
The stopping condition $f$ is true ($\top$) if the current refinement step $t$ reaches the number of refinement steps $T$ to be made.

The interact function $i$ exchanges the information which neighbor (0th, 1st, or 2nd) the particle $p_j$ is from the perspective of particle $p_k$ and saves this information in the $\gamma_j$ corresponding to the position of the index $i_k$ of the particle $p_k$ in $\beta_j$.
The evolve function $e$ creates the new particles  and uses the information stored in  $\gamma_j$ to determine the indices of the new particles.
The old particle does not exist anymore in the next state (refined triangulation), instead, a tuple of new particles  is created.
In this evolve function, the global variable $g$ remains unchanged.
The evolve method of the global variable $\mathring e$ increments the refinement step $t$ by $1$.

An instance of this particle method has to specify the initial set of triangles (particles) and the number of refinement steps that are to be computed. We recapitulate the situation in Fig.~\ref{fig:triangulation}, performing one refinement step of a single triangle into four smaller triangles. 
The method therefore starts from one particle with index $0$ and vertices $\left( (0,0) , (4,0), (2,4)\right)^T$.
The neighbor indices are all set to ${}^{r}\beta_1=-1$ ($r=0,1,2$) to indicate that there are no neighbors yet.
The reverse indices are arbitrarily set to ${}^{r}\gamma_1=0$ ($r=0,1,2$) but this has no meaning because there are no neighbors yet. This leads to the instance:
\begin{align*}
g^1:=&\left( t, T \right) &&\text{initial global variable}\\
T &:=1 &&\text{number of refinement steps}\\
t&:=0 &&\text{current refinement step}\\
\\
\mathbf p^1&:= (p^1_1) &&\text{initial particle}\\
p^1_1=&\left(\iota_1, \left(
\begin{array}{c}
{}^{0}v_1\\
{}^{1}v_1\\
{}^{2}v_1\\
\end{array}
 \right),  
 \left(
\begin{array}{c}
{}^{0}\beta_1\\
{}^{1}\beta_1\\
{}^{2}\beta_1\\
\end{array}
 \right),
 \left(
\begin{array}{c}
{}^{0}\gamma_1\\
{}^{1}\gamma_1\\
{}^{2}\gamma_1\\
\end{array}
 \right) \right) &&\text{initial triangle}\\
 :=&\left(0, \left(
\begin{array}{c}
(0 , 0)\\
(4 , 0)\\
(2 , 4)\\
\end{array}
 \right),  
 \left(
\begin{array}{c}
-1\\
-1\\
-1\\
\end{array}
 \right),
 \left(
\begin{array}{c}
0\\
0\\
0\\
\end{array}
 \right) \right)
\end{align*}
The result of executing this is as shown in Fig.~\ref{fig:triangulation}.

\subsection{Gaussian Elimination} \label{sec:example:gaus:example}
As a second example of a non-canonical particle method, we formulate the classic Gaussian elimination algorithm in the framework of our definition. Gaussian elimination is a classic algorithm to invert a matrix or solve a linear system of equations. It requires a cubic number of computation operations with linear dimension of the matrix.

In our example, we use Gaussian elimination without pivoting to solve a system of 3 linear equations with 3 unknowns. Each row of the resulting $3\times 3$ matrix represents the coefficients (weights of the linear combination) of one equation. The right-hand side values are stored in a 3-vector concatenated to the right of the matrix as shown below. 

Gaussian elimination uses three operations: swapping two rows, multiplying all entries in a row with a non-zero number, and adding one row to another.
These operations are combined in the following steps to reduce the matrix to reduced-row echelon form, i.e., an upper triangular matrix where the first non-zero entry in each row is 1 and no other entry in any leading-1 column is non-zero:
\begin{enumerate}
\item Start with the entry in the first row and the first column.
\item If the entry is 0, find the first row with a non-zero entry in the same column and swaps the rows.
If there is no non-zero entry, proceed to the next column but stay in the same row and start this step over.
If the entry is non-zero, make all other entries below it in the same column zero by adding corresponding multiples of the row to the other rows.
Finally, multiple the row with a non-zero number such that its leading element becomes 1.
\item Proceed to the next row and column and repeat step 2.
This results in an upper triangular matrix with leading ones once arrived at the last column.
\item Finally, start from the bottom row and add  multiples of the current row to the other rows to get 0 entries in all columns containing a leading 1.
This results in the reduced-row echelon form of the linear equation system.
\end{enumerate}

In our particle method implementation of this algorithm, a particle represents a row (equation) of the linear equation system. Interactions between particles (rows) change the entries in the rows such as to create the upper triangular matrix and then the reduced-row echelon form.
The evolve function reduces normalizes the rows to having leading ones. In the language of our formal definition, this can be expressed as:
\begin{align}
g:=&\left( N , m, n \right) \quad \text{ for } g \in  \mathbb{N} \times \mathbb{N} \times \mathbb{N}, 	\\
p:=&\left( \left(\; {}^{l}a\right)_{l=1}^{N}, b, \mu \right) \quad \text{ for } p \in P := \mathbb{R}^N \times \mathbb{R} \times \mathbb{R}, 	\\
\notag\\
u \left( [g, \mathbf p],  j \right):=&
\begin{cases}
     (N, ... , n+1)& \text{if} \quad j=n,\; m\leq N\\
     (1, ... , n-1)& \text{if} \quad  j=n,\; m> N  \\
      ()& \text{else, } 
   \end{cases}
   \\
f \left(\boldsymbol g \right):=& \left( n=1 \land m > N\right),\\
i(g, p_j, p_k):=& 
\begin{cases}
     \left(p_j, \left( \left( \;{}^{l}a_k-  \; {}^{l}a_j \; \frac{ {}^{m}a_k}{{}^{m}a_j}\right)_{l=m}^{N}, \; b_k- b_j\; \frac{  \: {}^{m}a_k}{ \: {}^{m}a_j}, \; \mu_k \right)\right)& \text{if} \quad m\leq N,\; {}^{m}a_j \neq 0\\
     (p_k, p_j)&  \text{if} \quad m\leq N,\; {}^{m}a_j = 0,\;  {}^{m}a_k\neq 0\\
      (p_j, p_k)& \text{if} \quad m\leq N,\; {}^{m}a_j = 0,\;  {}^{m}a_k = 0\\
      \left(p_j, \left( \left( {}^{l}a_k-  {}^{\mu_j}a_k \; {}^{l}\!a_j \right)_{l=\mu_j}^{N}, \; b_k- {}^{\mu_j}a_k\; b_j,\;  \mu_k \right)\right)& \text{else,}
   \end{cases}
   \\[12pt]
   \begin{split}
       e(g, p_j):=& \left( \vphantom{\begin{cases} \\[45pt] \end{cases}}
       \hspace{2mm}  \left(  N, m, 
        \begin{cases}
        n+1 & \text{if} \quad j=n, \;m<N,\; {}^{m}a_j \neq 0\\
        n-1  & \text{if} \quad j=n, \;m=N,\; {}^{m}a_j = 0\\
           n  & \text{else}
       \end{cases} \right) \right. ,\\
    &\left. \left(
       \begin{cases}
         \left( \left(\frac{{}^{l}a_j}{{}^{m}a_j} \right)_{l=m}^{N}, \frac{ b_j}{{}^{m}a_j}, m \right)& \text{if} \quad j=n, \;m\leq N,\; {}^{m}a_j \neq 0\\
               p_j & \text{else}
       \end{cases} 
       \right) \hspace{2mm} \vphantom{\begin{cases} \\[45pt] \end{cases}}\right) ,
    \end{split}
   \\[12pt]
   \mathring e(g):=& \left(N, m+1,
    \begin{cases}
       n-1 & \text{if} \quad m > N\\
       n  & \text{else}
   \end{cases}
    \right).
\end{align}

Here, the global variable $g$ is a collection of three natural numbers: the number of rows and columns $N$ of the square matrix, the index of the current column $m$ ($m>N$ indicates that the algorithm is in step 4), and the index of the current row $n$.
Each particle stores the matrix entries in the row represented by the particle in a vector $\left(\; {}^{l}a\right)_{l=1}^{N}$ with $b$ the corresponding right-hand side of the linear equation system.
For simplicity, the column index of the leading 1 is saved in a separate property $\mu$.

The neighborhood function $u$ ensures that only the current row ($j=n$) has neighbors, and therefore only this one row will participate in the pivoting process.
As long as $m\leq N$, the neighbor particles are the rows below in reverse order.
The reverse order ensures that if there is a zero row it will be sorted to the end by the interaction function $i$.
For step 4 of the algorithm ($m> N$), the order of the neighbor particles is irrelevant as the neighbor particles are then all the rows above.
The stopping condition $f$ is true ($\top$) if the current row is the top row again ($n=1$) and the algorithm is in step 4 ($m>N$).

The first three cases in the interact function $i$ are for steps 1 to 3 of the algorithm.
The last case for step 4.
The first case is the addition of a suitable multiple of the  current row to a neighboring row to create zeros in the $m$-th column.
The second case handles the swapping of two rows if the $m$-th entry of the current row $p_j$ is zero but that of the neighbor row $p_k$ is non-zero.
If the $m$-th entry of both interacting rows, $p_j$ and $p_k$, is zero then nothing happens (third case).
The fourth case creates zeros in the $\mu_j$-th (leading 1 entry) column of the interacting rows $p_k$.

The evolve function $e$ is only active as long as $m\leq N$ and only if the evolving particle is the current row of  $j=n$. Otherwise, it amounts to the identity map. 
In the non-trivial case, it normalizes the leading entry  ${}^{m}a_j$ of the current particle (row) to 1 by dividing all non-zero coefficients with ${}^{m}a_j$.
It also saves the index of the leading 1, $m$, in the particle property $\mu_j$.
The global variable $g$ is changed, incrementing the index of the current row $n$ if the current column is not the last one ($m<N$) and if the $m$-th coefficient of the row is not zero (${}^{m}a_j \neq 0$).
The turning point of the algorithm, where it switches from iterating downward through the rows to iterating back up again, is when $m=N$.
At this point, the index of the current row $n$ remains unchanged unless the last entry is zero (${}^{m}a_j = 0$), because then the row is completely zero and hence has no leading one.
In this case, the algorithm can already go to the row above ($n-1$).

The evolve function of the global variable $\mathring e$ increments the current column $m$ by one, so that the algorithm proceeds one column to the right in each iteration.
During step 4, $m$ increments beyond $N$, but this is irrelevant.
Then, however, $n$ is decremented by one in each iteration in order to move up one row at a time.

An instance of this particle method is defined by specifying the coefficients of the linear system as initial set of particles and the size of the system. The property $\mu$ can be initialized arbitrarily, as it is overwritten anyway before being used. As an example, we consider the instance for the $3\times 3$ equation system:
$$
\left[{ \begin{array}{rrr}
1&2&5\\
1&-1&-4\\
2&6&16
\end{array} }\right] \vec{x} = \left[ \begin{array}{r} 2 \\ -4 \\ 8 \end{array}\right]
$$
with unknown variable $\vec{x}$. For this example, $N=3$ and the particle method instance is:
\begin{align*}
g:=&\left( N , m, n \right) &&\text{initial global variable}\\
N:=&3 &&\text{number of rows and columns}\\
m:=&1 &&\text{start from the first column}\\
n:=&1 &&\text{start from the first row}\\
\\
\mathbf p^1 :=& (p_1,  p_2, p_3) &&\text{initial particle tuple}\\
p:=&( ( {}^{0}a, \;   {}^{1}a,\; {}^{2}a ), b, \mu ) &&\text{particle prototype}\\
p_1:=&\left( ( 1, \;\; 2 , \;\; 5), \;2, \;0 \right) &&\text{particle 1 (first row)}\\
p_2:=&\left( ( 1, -1 , -4), -4, 0 \right) &&\text{particle 2 (second row)}\\
p_3:=&\left( ( 2, \;\; 6 ,\; 16), \;8, \;0 \right) &&\text{particle 3 (third row)}\\
\end{align*}

Executing the algorithm for this instance produces the four iterations:
\begin{align*}
\left[{ \begin{array}{rrr|r}
1&2&5&2\\
1&-1&-4&-4\\
2&6&16&8
\end{array} }\right] 
\to 
\left[{\begin{array}{rrr|r}
1&2&5&2\\
0&-3&-9&-6\\
0&2&6&4
\end{array}}\right]
\to
\left[{\begin{array}{rrr|r}
1&2&5&2\\
0&-3&-9&-6\\
0&0&0&0
\end{array}}\right]\\
\\
\to 
\left[{\begin{array}{rrr|r}
1&2&5&2\\
0&1&3&2\\
0&0&0&0
\end{array}}\right]
\to 
\left[{\begin{array}{rrr|r}
1&0&-1&-2\\
0&1&3&2\\
0&0&0&0
\end{array}}\right].
\end{align*}

\section{Conclusion and Discussion} \label{sec:conclusion}

We have presented a formal definition of the algorithmic class of particle methods. While particle methods have traditionally been used in plasma physics \cite{Hockney:1966} and in computational fluid dynamics \cite{Chorin:1973,Cottet:1990}, our definition equally applies to popular algorithms in image processing \cite{Cardinale:2012, Afshar:2016}, computer graphics \cite{Gross:2011uz}, and computational optimization~\cite{Hansen:1996, Muller:2010}. The proposed definition therefore highlights the algorithmic commonalities in the computational structure across applications, and it enables a sharp classification of what constitutes a particle method and what does not. The present definition unifies all particle methods and also enables formulating particle methods for non-canonical problems that were so far solved differently.

The core of our definition is that particles are only allowed to interact with one another pairwise. 
We illustrated the application of this definition in several examples, covering both canonical particle methods and non-canonical ones. As examples of canonical particle methods, we presented a particle strength exchange (PSE, \cite{Degond:1989a})  simulation of the continuum diffusion equation and a Lennard-Jones \cite{Lennard-Jones:1931} molecular dynamics simulation of a discrete model. As non-canonical examples, we presented how triangulation refinement and Gaussian elimination can be formulated as particle methods. While this illustrates the unifying nature of the presented definition, it does not imply that such an implementation would be beneficial in any way. 
However, it prospectively enables theoretical analysis, comparison, and hopefully also the combination of  algorithms from different fields under a common framework.

We deliberately formulated the present definition in the most general way in order for it to encompass everything that we call a particle method. Most practical instances, however, do not exploit the full generality of the definition. For example, one would frequently restrict a particle method to be order-independent, i.e., to produce results that are independent of the indexing order of the particles. Another common restriction is to limit the neighborhood function to only be able to access particles within a certain radius, e.g. using cell list \cite{Hockney:1988} or Verlet list \cite{Verlet:1967} algorithms, or to restrict the computational power of the interact and evolve functions or the memory of the particle state.

Such restrictions are also frequently needed in order to parallelize particle methods on multi-processor computer architectures. 
While our definition is general, it uses sequential state transitions.
A definition of concurrent or parallel particle methods could be achieved by additional restriction on our particle methods definition. In the examples presented, we observe that parallelization is straightforward in all cases, except for Gaussian elimination. Gaussian elimination, when formulated as a particle method, is inherently sequential and the dynamically growing particle size would hamper parallel efficiency and load balance. 
The generality of our definition therefore prevents efficient concurrent formulations of particle methods unless it is suitably restricted. 

Another limitation of the presented definition of particle methods is its monolithic nature.
An algorithm composed of many smaller algorithms, like for example a solver for the incompressible Navier-Stokes equation, will become very large and complex with several nested cases when explicitly formulated in our definition. Finally, our definition is not unique. Alternative, possibly more compact or elegant, but equivalent definitions are possible. We chose the formulation presented here, as we believe it to be close to practical implementations.

Notwithstanding these limitations, the present definition establishes the so-far loose notion of particle methods as a rigorous algorithmic class. This paves the way for future research both in the theoretical algorithmic foundations of particle methods as well as in the engineering of their software implementation.

Future theoretical work could, for example, define different classes of particle methods by formalizing class-specific restrictions to our definition. This would enable classifying particle methods with respect to their implementation complexity, degree of concurrency, or computational power.
Different classes of particle methods could correspond to different implementation scenarios, or to different requirements, such as order-independence.
Other possible directions of theoretical research include the derivation of complexity bounds for certain classes of particle methods.
While  it seems intuitive that the present definition of particle methods is Turing-powerful (because one could use a single particle to implement a universal Turing machine in the evolve method), this trivial reduction offers no insight into the algorithmic structure and its computational power. 
Studying the computational power of certain classes of particle methods could therefore provide interesting insight into what is possible with different classes of particle methods.

On the engineering side, future work can leverage the present definition to better structure software frameworks for particle methods, auch as the PPM Library \cite{Sbalzarini:2006b}, OpenFPM \cite{Incardona:2019}, POOMA \cite{Reynders:1996a}, or FDPS \cite{Iwasawa:2016}. This would render them easier to understand to users and more maintainable, as the formal definition provides a common vocabulary. The present definition also enables classifications of software frameworks with respect to their expressiveness, coverage of the definition, or optimization toward specific classes of particle methods.    

Future work could also attempt a less monolithic definition that allows modular combinations of different particle methods. While this could lead to a formulation that can potentially directly be exploited in software engineering or in the design of domain-specific programming languages for particle methods \cite{Karol:2018}, it does require solving a few theoretical problems: One such problem is how to synchronize data when different particle methods access shared particles. Other open issues include how different types of particles from different methods can interact with one another (e.g., for remeshing) and how access of an algorithm can be restricted to a subset of  particles (e.g., for boundary element methods). All of these need to be formally described first, which might necessitate additional data structures or functions to be added to the present definition.

Taken together, the present formal definition of particle methods is a necessary first step towards a sound and formal understanding of what particle methods are, what they can do, and how efficient and powerful they are. 
It also provides guidance for practical software implementations and enables their comparative evaluation on common grounds. We therefore hope that the present work will generate downstream results and studies in different fields of application of particle methods.

\section*{Acknowledgments}
We thank Udo Hebisch (TU Bergakademie Freiberg, Germany) for discussions and comments on an early version of this manuscript. We thank Georges-Henri Cottet, Jean-Baptiste Keck, Christophe Picard, Aude Maignan, Cl\'{e}ment Pernet (Universit\'{e} Grenoble Alpes, France) for discussions on the application examples. We thank Ulrik G\"{u}nther for proofreading. This work has been funded by the German Research Foundation (Deutsche Forschungsgemeinschaft, DFG) within the Research Training Group ``Role-based Software Infrastructures for continuous-context-sensitive Systems''  (GRK 1907).

\bibliographystyle{plain}


\begin{thebibliography}{10}

\bibitem{Afshar:2016}
Yaser Afshar and Ivo~F. Sbalzarini.
\newblock A parallel distributed-memory particle method enables
  acquisition-rate segmentation of large fluorescence microscopy images.
\newblock {\em PLoS One}, 11(4):e0152528, 2016.

\bibitem{Alder:1957}
B~J Alder, TE~Wainwright J Chem~Phys Internet, and {1957}.
\newblock {Molecular dynamics simulation of hard sphere system}.

\bibitem{Bergdorf:2010}
Michael Bergdorf, Ivo~F. Sbalzarini, and Petros Koumoutsakos.
\newblock A {L}agrangian particle method for reaction-diffusion systems on
  deforming surfaces.
\newblock {\em J.\ Math.\ Biol.}, 61:649--663, 2010.

\bibitem{Bourantas:2016}
George~C. Bourantas, Bevan~L. Cheeseman, Rajesh Ramaswamy, and Ivo~F.
  Sbalzarini.
\newblock Using {DC PSE} operator discretization in {E}ulerian meshless
  collocation methods improves their robustness in complex geometries.
\newblock {\em Computers \& Fluids}, 136:285--300, 2016.

\bibitem{Caprace:2020}
Denis-Gabriel Caprace, Gr{\'e}goire Winckelmans, and Philippe Chatelain.
\newblock {An immersed lifting and dragging line model for the vortex
  particle-mesh method}.
\newblock {\em Theoretical and Computational Fluid Dynamics}, 34(1):21--48,
  2020.

\bibitem{Cardinale:2012}
Janick Cardinale, Gr\'{e}gory Paul, and Ivo~F. Sbalzarini.
\newblock Discrete region competition for unknown numbers of connected regions.
\newblock {\em IEEE Trans. Image Process.}, 21(8):3531--3545, 2012.

\bibitem{Chorin:1973}
A.~J. Chorin.
\newblock Numerical study of slightly viscous flow.
\newblock {\em J.\ Fluid Mech.}, 57(4):785--796, 1973.

\bibitem{Cottet:1990}
G.~H. Cottet and S.~Mas-Gallic.
\newblock A particle method to solve the {N}avier-{S}tokes system.
\newblock {\em Numer.\ Math.}, 57:805--827, 1990.

\bibitem{cottet:2020}
GEORGES-HENRI COTTET.
\newblock Two dimensional incompressible fluid flow.
\newblock {\em Mathematical Topics in Fluid Mechanics}, page~32, 2020.

\bibitem{Cottet:2014}
Georges-Henri Cottet, Jean-Matthieu Etancelin, Franck P{\'e}rignon, and
  Christophe Picard.
\newblock High order semi-{L}agrangian particle methods for transport
  equations: numerical analysis and implementation issues.
\newblock {\em ESAIM: Mathematical Modelling and Numerical Analysis},
  48(4):1029--1060, 2014.

\bibitem{Degond:1989a}
P.~Degond and S.~Mas-Gallic.
\newblock The weighted particle method for convection-diffusion equations.
  {P}art 1: {T}he case of an isotropic viscosity.
\newblock {\em Math.\ Comput.}, 53(188):485--507, 1989.

\bibitem{Degond:1989}
P.~Degond and S.~Mas-Gallic.
\newblock The weighted particle method for convection-diffusion equations.
  {P}art 2: {T}he anisotropic case.
\newblock {\em Math.\ Comput.}, 53(188):509--525, 1989.

\bibitem{Despreaux:2018}
S~Despr{\'e}aux, A~Maignan 2018 20th~International Symposium, and {2018}.
\newblock {GPaR: A Parallel Graph Rewriting Tool}.
\newblock {\em ieeexplore.ieee.org}, 2019.

\bibitem{Eldredge:2002}
Jeff~D. Eldredge, Anthony Leonard, and Tim Colonius.
\newblock A general deterministic treatment of derivatives in particle methods.
\newblock {\em J.\ Comput.\ Phys.}, 180:686--709, 2002.

\bibitem{Gingold:1977}
R.~A. Gingold and J.~J. Monaghan.
\newblock Smoothed particle hydrodynamics - theory and application to
  non-spherical stars.
\newblock {\em Royal Astronomical Society, Montly Notices}, 181:375--378, 1977.

\bibitem{Gross:2011uz}
Markus Gross and Hanspeter Pfister.
\newblock {\em {Point-Based Graphics}}.
\newblock Elsevier, May 2011.

\bibitem{Hansen:1996}
Nikolaus Hansen and Andreas Ostermeier.
\newblock {Adapting Arbitrary Normal Mutation Distributions in Evolution
  Strategies: The Covariance Matrix Adaptation}.
\newblock In {\em Proceedings of the 1996 IEEE Conference on Evolutionary
  Computation ({ICEC '96})}, pages 312--317, 1996.

\bibitem{Hockney:1966}
R.~W. Hockney.
\newblock Computer experiment of anomalous diffusion.
\newblock {\em Phys.\ Fluids}, 9(9):1826--1835, 1966.

\bibitem{Hockney:1988}
R.~W. Hockney and J.~W. Eastwood.
\newblock {\em Computer Simulation using Particles}.
\newblock Institute of Physics Publishing, 1988.

\bibitem{Incardona:2019}
Pietro Incardona, Antonio Leo, Yaroslav Zaluzhnyi, Rajesh Ramaswamy, and Ivo~F.
  Sbalzarini.
\newblock {OpenFPM}: A scalable open framework for particle and particle-mesh
  codes on parallel computers.
\newblock {\em Comput.\ Phys.\ Commun.}, 241:155--177, 2019.

\bibitem{Iwasawa:2016}
Masaki Iwasawa, Ataru Tanikawa, Natsuki Hosono, Keigo Nitadori, Takayuki
  Muranushi, and Junichiro Makino.
\newblock Implementation and performance of {FDPS}: a framework for developing
  parallel particle simulation codes.
\newblock {\em Publications of the Astronomical Society of Japan}, 68(4):54,
  2016.

\bibitem{Karol:2018}
Sven Karol, Tobias Nett, Jeronimo Castrillon, and Ivo~F. Sbalzarini.
\newblock A domain-specific language and editor for parallel particle methods.
\newblock {\em ACM Trans. Math. Softw.}, 44(3):34, 2018.

\bibitem{Kozen:1997}
Dexter~C. Kozen.
\newblock {\em Automata and Computability}.
\newblock Springer, 1997.

\bibitem{Lennard-Jones:1931}
J~E Lennard-Jones.
\newblock Cohesion.
\newblock {\em Proceedings of the Physical Society}, 43(5):461--482, sep 1931.

\bibitem{Liu:1995}
Wing~Kam Liu, Sukky Jun, and Yi~Fei Zhang.
\newblock Reproducing kernel particle methods.
\newblock {\em Int.\ J.\ Numer.\ Meth.\ Fluids}, 20:1081--1106, 1995.

\bibitem{SUMO2018}
Pablo~Alvarez Lopez, Michael Behrisch, Laura Bieker-Walz, Jakob Erdmann,
  Yun-Pang Fl{\"o}tter{\"o}d, Robert Hilbrich, Leonhard L{\"u}cken, Johannes
  Rummel, Peter Wagner, and Evamarie Wie{\ss}ner.
\newblock Microscopic traffic simulation using sumo.
\newblock In {\em The 21st IEEE International Conference on Intelligent
  Transportation Systems}. IEEE, 2018.

\bibitem{Monaghan:2005}
J.~J. Monaghan.
\newblock Smoothed particle hydrodynamics.
\newblock {\em Rep.\ Prog.\ Phys.}, 68:1703--1759, 2005.

\bibitem{Muller:2010}
Christian~L. M\"{u}ller and Ivo~F. Sbalzarini.
\newblock {G}aussian {A}daptation revisited --- an entropic view on covariance
  matrix adaptation.
\newblock In {\em Proc.\ EvoStar}, volume 6024 of {\em Lect. Notes Comput.
  Sci.}, pages 432--441, Istanbul, Turkey, April 2010. Springer.

\bibitem{Quentrec:1973}
B~Quentrec and C~Brot.
\newblock New method for searching for neighbors in molecular dynamics
  computations.
\newblock {\em Journal of Computational Physics}, 13(3):430 -- 432, 1973.

\bibitem{Reboux:2012}
Sylvain Reboux, Birte Schrader, and Ivo~F. Sbalzarini.
\newblock A self-organizing {L}agrangian particle method for
  adaptive-resolution advection--diffusion simulations.
\newblock {\em J.\ Comput.\ Phys.}, 231:3623--3646, 2012.

\bibitem{Reynders:1996a}
J.V.W. Reynders, J.C. Cummings, M.~Tholburn, P.J. Hinker, S.R. Atlas,
  S.~Banerjee, M.~Srikant, W.F. Humphrey, S.R. Karmesin, and K.~Keahey.
\newblock Pooma: a framework for scientific simulation on parallel
  architectures.
\newblock In A.~Bode, M.~Gerndt, R.G. Hackenberg, and H.~Hellwagner, editors,
  {\em Proceedings. First International Workshop on High-Level Programming
  Models and Supportive Environments}, pages 41--49, Los Alamitos, CA, USA,
  1996. Tech. Univ. Munchen; Res. Centre Julich; Central Inst. Appl. Math.;
  10th IEEE Int. Parallel Process. Symposium; IEEE Comput. Soc. Tech. Committee
  on Parallel Process.; ACM SIGARCH, {IEEE Comput. Soc. Press}.

\bibitem{Sbalzarini:2006b}
I.~F. Sbalzarini, J.~H. Walther, M.~Bergdorf, S.~E. Hieber, E.~M. Kotsalis, and
  P.~Koumoutsakos.
\newblock {PPM} -- a highly efficient parallel particle-mesh library for the
  simulation of continuum systems.
\newblock {\em J.\ Comput.\ Phys.}, 215(2):566--588, 2006.

\bibitem{Sbalzarini:2005}
Ivo~F. Sbalzarini, Anna Mezzacasa, Ari Helenius, and Petros Koumoutsakos.
\newblock Effects of organelle shape on fluorescence recovery after
  photobleaching.
\newblock {\em Biophys.\ J.}, 89(3):1482--1492, 2005.

\bibitem{Schrader:2010}
Birte Schrader, Sylvain Reboux, and Ivo~F. Sbalzarini.
\newblock Discretization correction of general integral {PSE} operators in
  particle methods.
\newblock {\em J.\ Comput.\ Phys.}, 229:4159--4182, 2010.

\bibitem{Silbert:2001}
Leonardo~E. Silbert, Deniz Erta\c{s}, Gary~S. Grest, Thomas~C. Halsey, Dov
  Levine, and Steven~J. Plimpton.
\newblock Granular flow down an inclined plane: {B}agnold scaling and rheology.
\newblock {\em Phys.\ Rev.\ {E}}, 64:051302, 2001.

\bibitem{Swope:1982}
William~C. Swope, Hans~C. Andersen, Peter~H. Berens, and Kent~R. Wilson.
\newblock A computer simulation method for the calculation of equilibrium
  constants for the formation of physical clusters of molecules: Application to
  small water clusters.
\newblock {\em The Journal of Chemical Physics}, 76(1):637--649, 1982.

\bibitem{Verlet:1967}
L.~Verlet.
\newblock Computer experiments on classical fluids. {I.} {T}hermodynamical
  properties of {L}ennard-{J}ones molecules.
\newblock {\em Phys.\ Rev.}, 159(1):98--103, 1967.

\bibitem{Walther:2009}
Jens~H. Walther and Ivo~F. Sbalzarini.
\newblock Large-scale parallel discrete element simulations of granular flow.
\newblock {\em Engineering Computations}, 26(6):688--697, 2009.

\end{thebibliography}

\end{document}